\documentclass{egpubl}
\usepackage{egsr2024}

\def\finalversion{1}

\if\finalversion1
\WsPaper           
\else
\WsSubmission      
\fi

\CGFccby

\usepackage[T1]{fontenc}
\usepackage{dfadobe}

\biberVersion
\BibtexOrBiblatex
\usepackage[backend=biber,bibstyle=EG,citestyle=alphabetic,backref=true]{biblatex}
\electronicVersion
\PrintedOrElectronic

\ifpdf \usepackage[pdftex]{graphicx} \pdfcompresslevel=9
\else \usepackage[dvips]{graphicx} \fi

\usepackage{egweblnk}
\usepackage{amsmath}
\usepackage{amssymb}
\usepackage{todonotes}
\usepackage{tikz}
\usepackage{pgfplots}
\usepackage{nicefrac}
\usepackage{bigints}
\usepackage{tabularx}
\usepackage{booktabs}
\usetikzlibrary{math}
\usetikzlibrary{calc}

\def\etal{\textit{et al.}}
\def\eg{\textit{e.g.}}
\def\ie{\textit{i.e.}}

\if\finalversion1
\newcommand\aftodo[1]{}

\newcommand\laurent[1]{}
\newcommand\edit[1]{{#1}}
\else
\newcommand\aftodo[1]{{\color{blue}TODO: #1}}

\newcommand\laurent[1]{{\color{blue}{\textbf{[Laurent:} #1\textbf{]}}}}
\newcommand\edit[1]{{\color{purple}#1}}
\fi

\newcommand{\li}{\lambda_i}
\newcommand{\lo}{\lambda_o}
\newcommand{\wi}{\omega_i}
\newcommand{\wo}{\omega_o}
\newcommand{\wn}{\omega_n}

\newcommand{\rerad}{\mathcal{P}}
\newcommand{\redux}{\mathrm{P}}
\newcommand{\reduxh}{\redux^{\mbox{\scriptsize Naive}}}
\newcommand{\reduxo}{\redux^{\mbox{\scriptsize Ours}}}
\newcommand{\down}{\mbox{\textbf{down}}}
\newcommand{\up}{\mbox{\textbf{up}}}

\newcommand{\dd}[1]{\mbox{d}#1}

\definecolor{tab_blue}{rgb}{0.12156862745098039,0.4666666666666667,0.7058823529411765}
\definecolor{tab_orange}{rgb}{1.0,0.4980392156862745,0.054901960784313725}
\definecolor{tab_green}{rgb}{0.17254901960784313,0.6274509803921569,0.17254901960784313}
\definecolor{tab_red}{rgb}{0.8392156862745098,0.15294117647058825,0.1568627450980392}
\definecolor{tab_purple}{rgb}{0.5803921568627451,0.403921568627451,0.7411764705882353}
\definecolor{tab_brown}{rgb}{0.5490196078431373,0.33725490196078434,0.29411764705882354}
\definecolor{tab_pink}{rgb}{0.8901960784313725,0.4666666666666667,0.7607843137254902}
\definecolor{tab_gray}{rgb}{0.4980392156862745,0.4980392156862745,0.4980392156862745}
\definecolor{tab_olive}{rgb}{0.7372549019607844,0.7411764705882353,0.13333333333333333}
\definecolor{tab_cyan}{rgb}{0.09019607843137255,0.7450980392156863,0.8117647058823529}

\title[Non-Orthogonal Reduction for Rendering Fluorescent Materials in Non-Spectral Engines]%
      {Non-Orthogonal Reduction for Rendering Fluorescent Materials \\ in Non-Spectral Engines}

\if\finalversion1
\author[A. Fichet \etal]
{
\parbox{\textwidth}{    \centering %
    A. Fichet$^{1,\star}$\orcid{0000-0001-7756-0901},~%
    L. Belcour$^{1,\star}$\orcid{0000-0001-5923-423X},~and~%
    P. Barla$^{2}$\orcid{0000-0001-7756-0901}%
}
        \\
{\parbox{\textwidth}{\centering $^1$Intel Corporation, France, \;%
         $^2$Inria, France%
       }
}
}
\else
\author[paper1011]{paper1011}
\fi

\begin{document}

\teaser{
    \vspace{-1cm}
    \center
    \pgfdeclarelayer{background}
\pgfdeclarelayer{foreground}
\pgfsetlayers{background,main,foreground}
\begin{tikzpicture}[font=\footnotesize]
    \tikzmath{
        \teaserMargins        = 3;
        \pathLabelsWidth      = 10;
        \teaserHeight         = .22*\the\linewidth; 
        \heightPath           = (\teaserHeight - \teaserMargins) / 2;
        \heightPathWithMargin = \heightPath + \teaserMargins;
        \widthPath            = \heightPath;
        \widthPathWithMargin  = \heightPathWithMargin;
        \widthRealtime        = \the\linewidth - 3*\widthPathWithMargin - \pathLabelsWidth;
        \xStartPath           = \widthRealtime + \teaserMargins;
        \xStartLabels         = \xStartPath + 3*\widthPathWithMargin - \teaserMargins;
        \spitOffset           = .05;
        \splitTranslate       = .04/2 * \widthRealtime;
        \splitDayTop          = (.18 - \spitOffset)*\widthRealtime + \splitTranslate;
        \splitDayBottom       = (.18 + \spitOffset)*\widthRealtime + \splitTranslate;
        \splitBlueTop         = (.18 - \spitOffset)*\widthRealtime + .5*\widthRealtime + \splitTranslate;
        \splitBlueBottom      = (.18 + \spitOffset)*\widthRealtime + .5*\widthRealtime + \splitTranslate;
    }

    \begin{scope}
        \clip (0, 0) rectangle (.5*\widthRealtime pt, \teaserHeight pt);
        \node[anchor=south west, inner sep=0] (unity_day) at (0, 0) {\includegraphics[height=\teaserHeight pt]{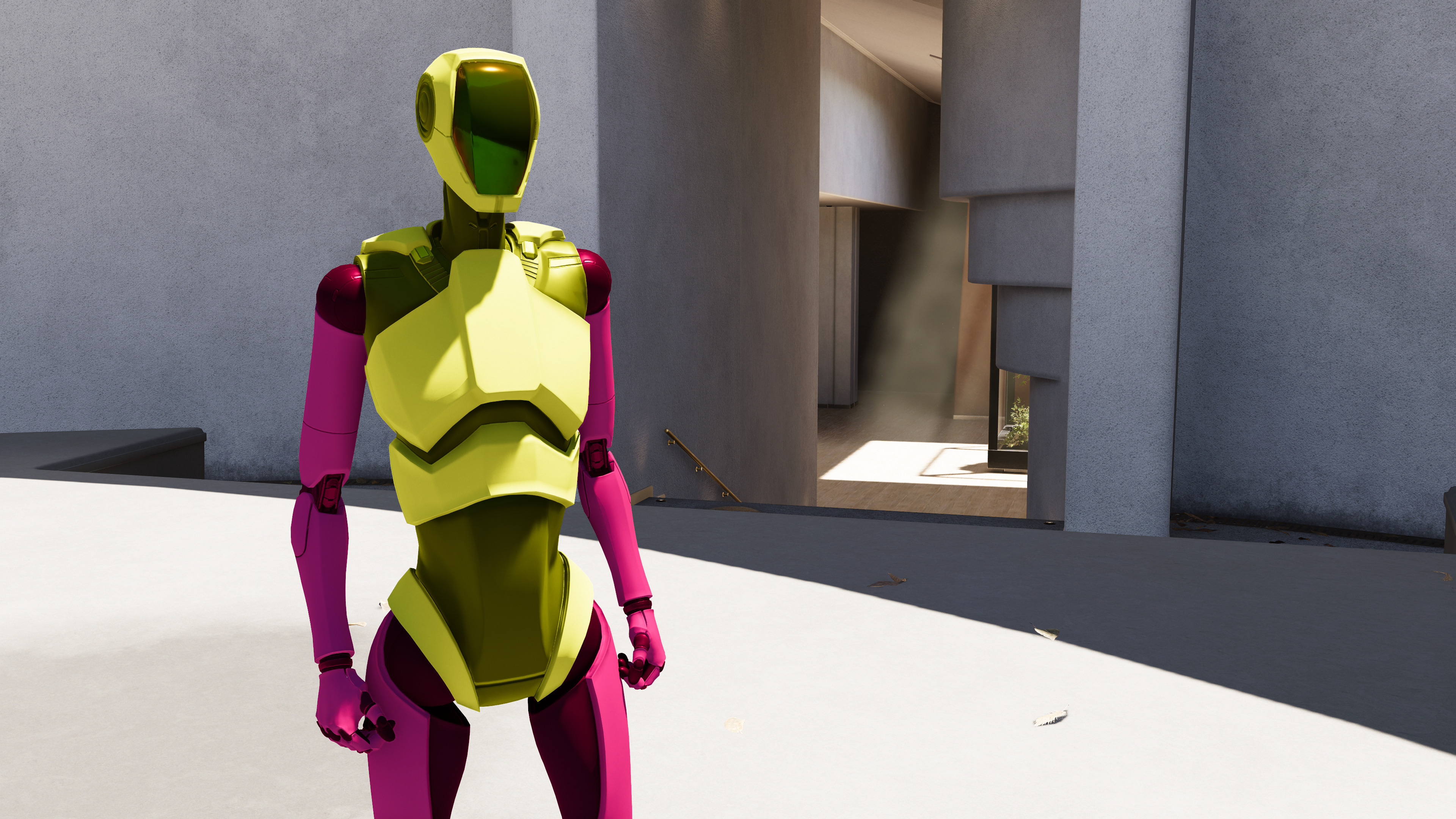}};
        \clip (\splitDayBottom pt, 0) -- (\splitDayTop pt, \teaserHeight pt) -- (\widthRealtime pt, \teaserHeight pt) -- (\widthRealtime pt, 0) -- cycle;
        \node[anchor=south west, inner sep=0] (unity_day) at (0, 0) {\includegraphics[height=\teaserHeight pt]{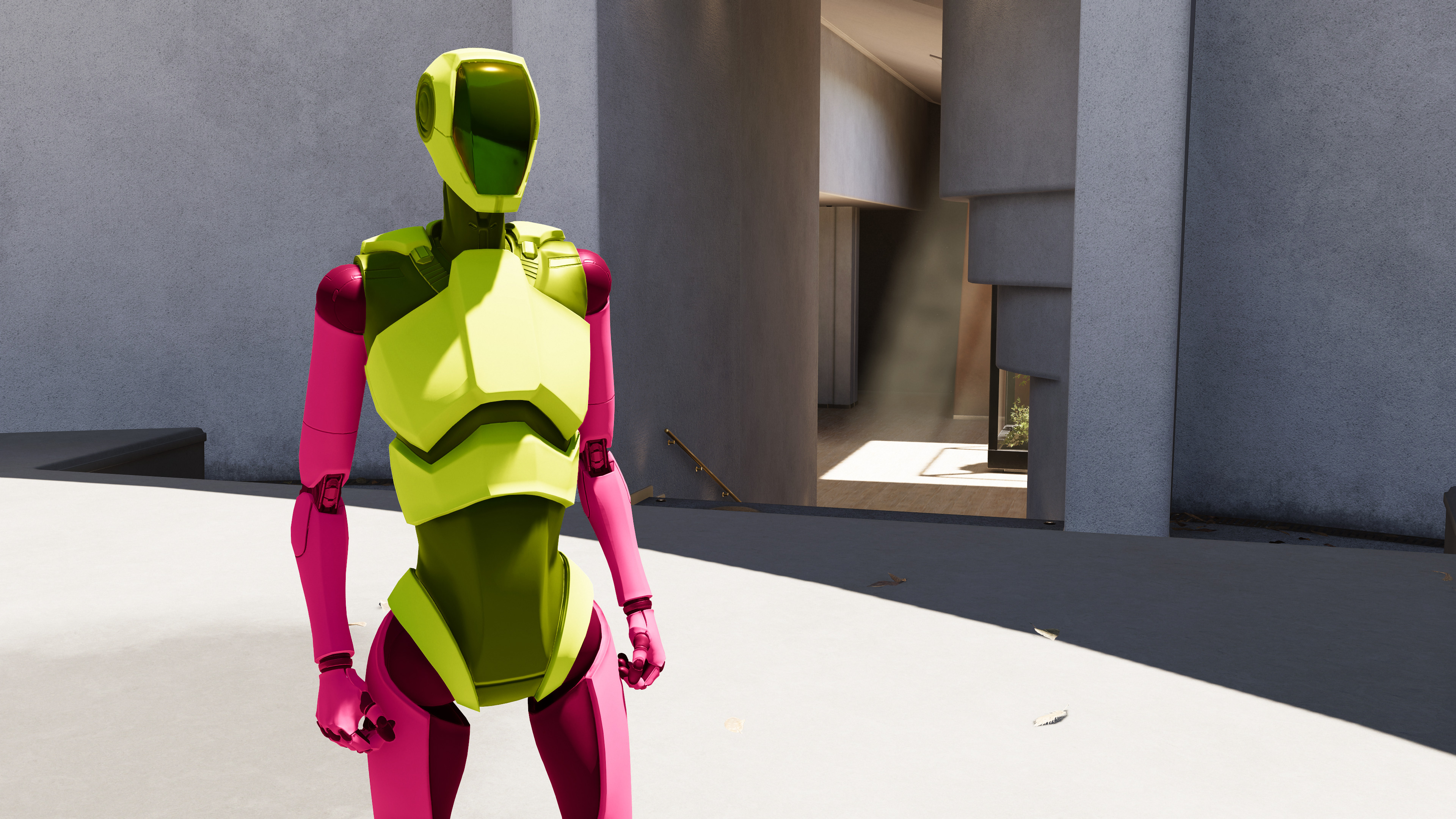}};
    \end{scope}
    \begin{scope}

        \clip (.5*\widthRealtime pt, 0) rectangle (\widthRealtime pt, \teaserHeight pt);
        \node[anchor=south west, inner sep=0] (unity_disco) at (.5*\widthRealtime pt, 0) {\includegraphics[height=\teaserHeight pt]{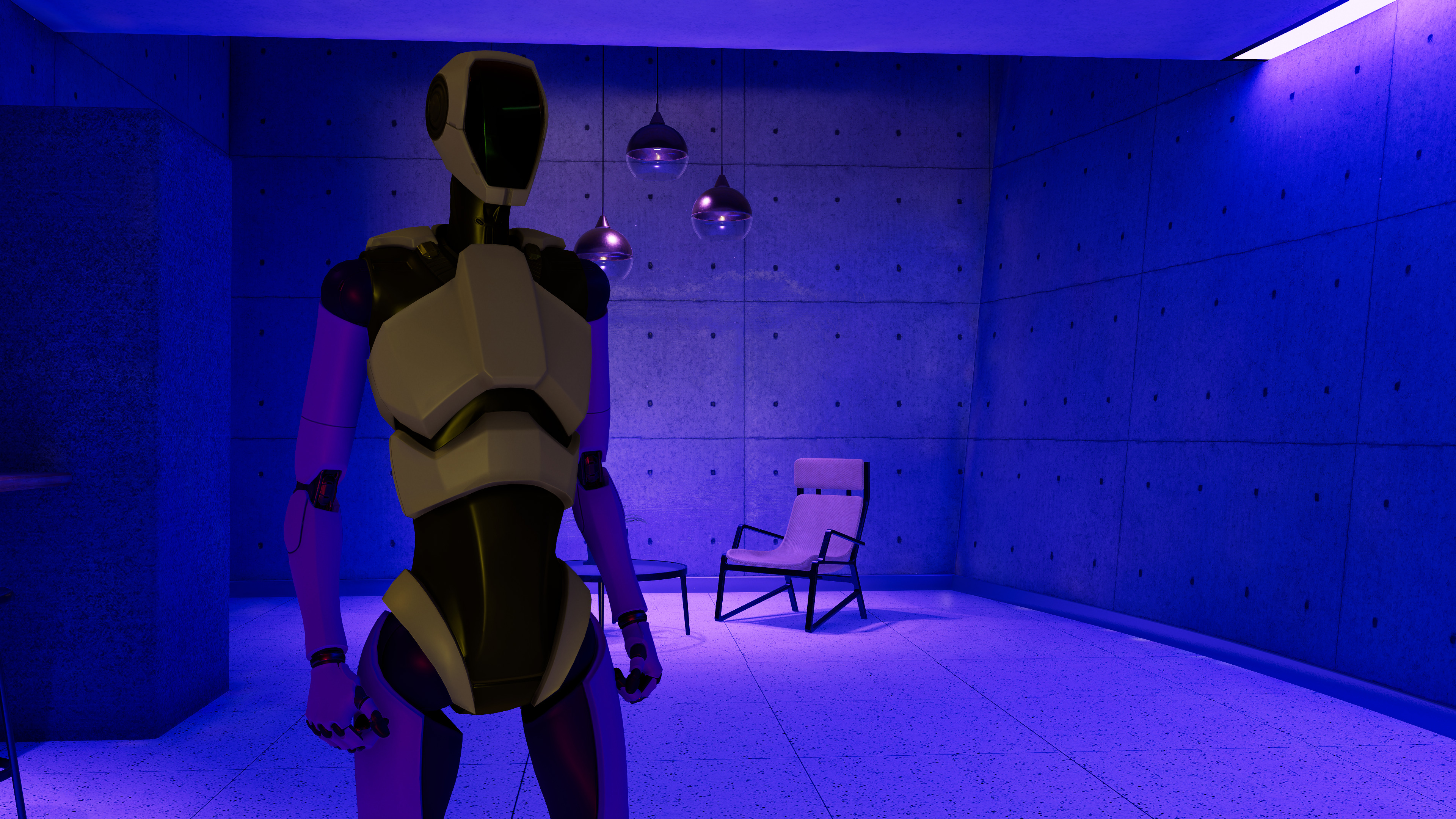}};
        \clip (\splitBlueBottom pt, 0) -- (\splitBlueTop pt, \teaserHeight pt) -- (\widthRealtime pt, \teaserHeight pt) -- (\widthRealtime pt, 0) -- cycle;
        \node[anchor=south west, inner sep=0] (unity_disco) at (.5*\widthRealtime pt, 0) {\includegraphics[height=\teaserHeight pt]{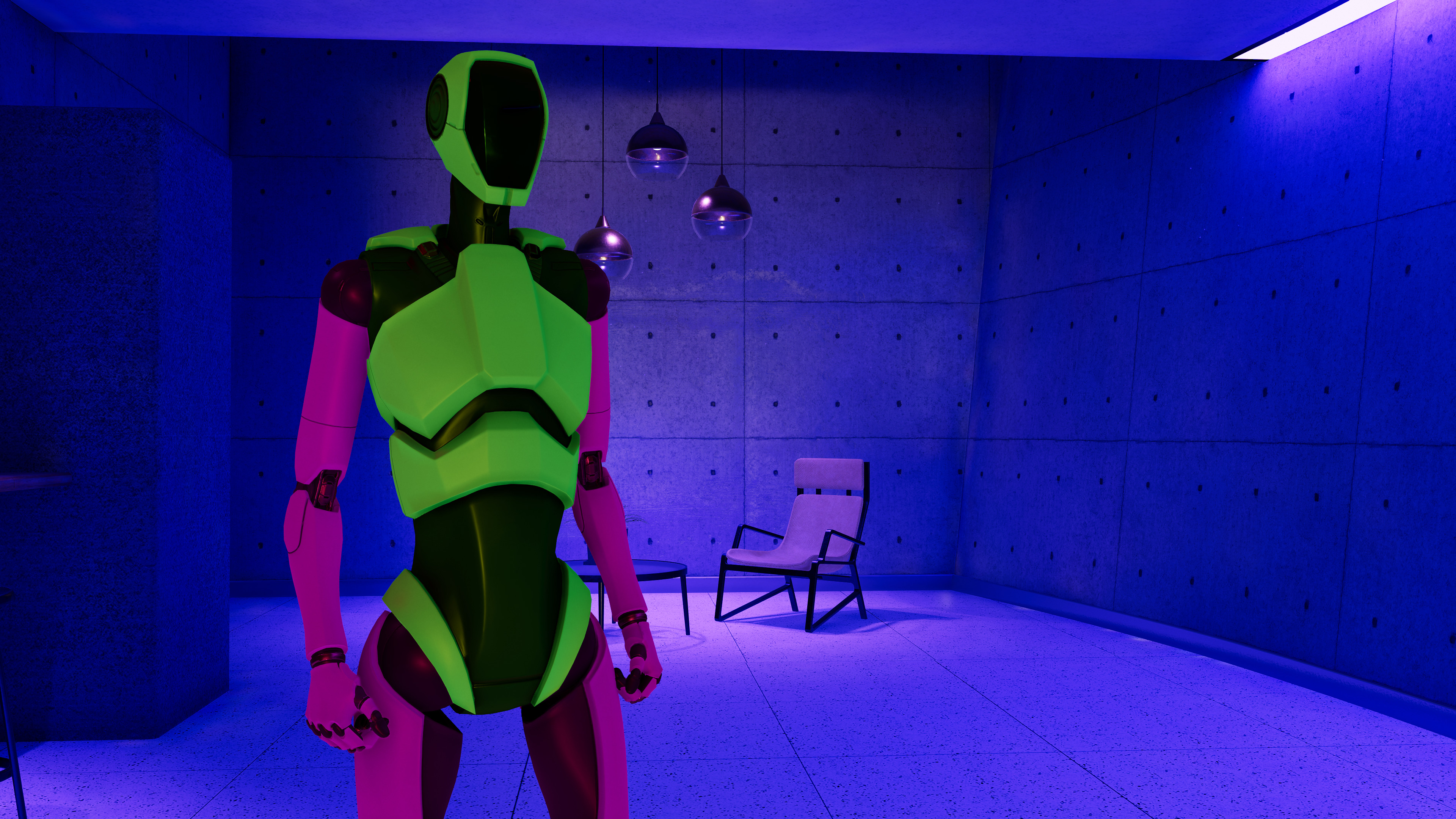}};
    \end{scope}

    \draw (0, 0) rectangle (\widthRealtime pt, \teaserHeight pt);
    \draw[line width=0.5pt] (.5*\widthRealtime pt, 0) -- (.5*\widthRealtime pt, \teaserHeight pt);

    \draw[line width=0.1pt] (\splitDayBottom pt, 0) -- (\splitDayTop pt, \teaserHeight pt);
    \draw[line width=0.1pt] (\splitBlueBottom pt, 0) -- (\splitBlueTop pt, \teaserHeight pt);

    \begin{pgfonlayer}{foreground}
        \node[anchor=base, yshift=3.6pt, xshift=.4pt] (txt) at (.5*\widthRealtime pt, -13 pt) {{(a) Real time rendering with our reduction}};
    \end{pgfonlayer}


    \node[above, anchor=south west, xshift=.4pt, yshift=-.4pt, text opacity=.8] at (0                   , 0) {\textcolor{black} {\tiny{Daylight}}};
    \node[above, anchor=south west                           , text opacity=.8] at (0                   , 0) {\textcolor{gray}{\tiny{Daylight}}};
    \node[above, anchor=south east, xshift=.4pt, yshift=-.4pt, text opacity=.8] at (\widthRealtime pt, 0) {\textcolor{gray} {\tiny{Blue light}}};
    \node[above, anchor=south east                           , text opacity=.8] at (\widthRealtime pt, 0) {\textcolor{white}{\tiny{Blue light}}};

    \node[anchor=north west, xshift=.4pt, yshift=-.4pt, text opacity=.8] at (0                   , \teaserHeight pt) {\textcolor{gray} {Fluo off}};
    \node[anchor=north west                           , text opacity=.8] at (0                   , \teaserHeight pt) {\textcolor{red}{Fluo off}};
    \node[anchor=north east, xshift=.4pt, yshift=-.4pt, text opacity=.8] at (.5*\widthRealtime pt, \teaserHeight pt) {\textcolor{gray} {Fluo on}};
    \node[anchor=north east                           , text opacity=.8] at (.5*\widthRealtime pt, \teaserHeight pt) {\textcolor{green}{Fluo on}};
    \node[anchor=north west, xshift=.4pt, yshift=-.4pt, text opacity=.8] at (.5*\widthRealtime pt, \teaserHeight pt) {\textcolor{gray} {Fluo off}};
    \node[anchor=north west                           , text opacity=.8] at (.5*\widthRealtime pt, \teaserHeight pt) {\textcolor{red}{Fluo off}};
    \node[anchor=north east, xshift=.4pt, yshift=-.4pt, text opacity=.8] at (   \widthRealtime pt, \teaserHeight pt) {\textcolor{gray} {Fluo on}};
    \node[anchor=north east                           , text opacity=.8] at (   \widthRealtime pt, \teaserHeight pt) {\textcolor{green}{Fluo on}};

    \begin{scope}[xshift=\xStartPath pt]
        \begin{scope}
            \clip (0, 0) rectangle +(.5*\widthPath pt, \teaserHeight pt);
            \node[anchor=south west, inner sep=0] (ixcclye_path_ours_3)  at (0, \heightPathWithMargin pt) {\includegraphics[height=\heightPath pt]{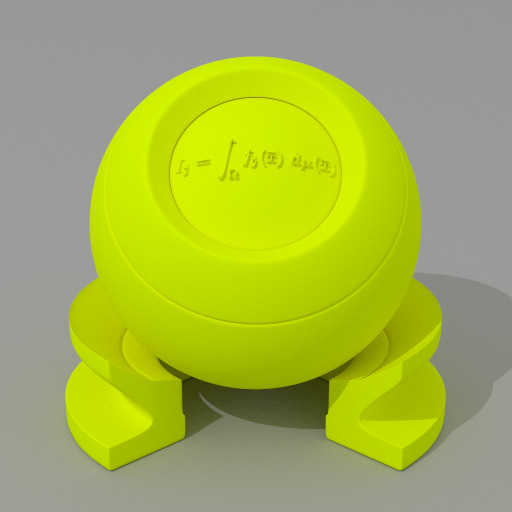}};
            \node[anchor=south west, inner sep=0] (herpipin_path_ours_3) at (0, 0                       ) {\includegraphics[height=\heightPath pt]{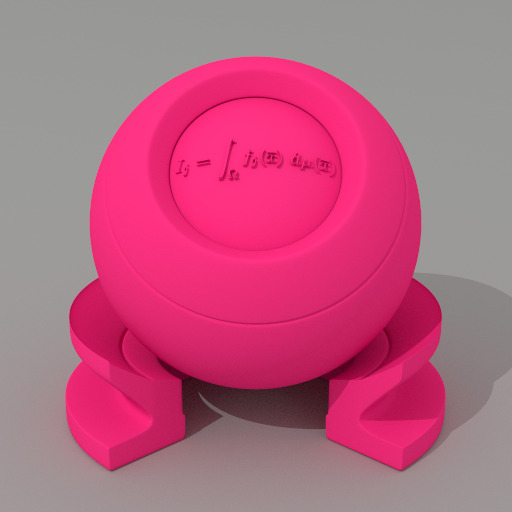}};
        \end{scope}

        \begin{scope}
            \clip[xshift=.5*\widthPath pt] (0 pt, 0 pt) rectangle +(.5*\widthPath pt, \teaserHeight pt);
            \node[anchor=south west, inner sep=0] (ixcclye_path_ours_3)  at (0, \heightPathWithMargin pt) {\includegraphics[height=\heightPath pt]{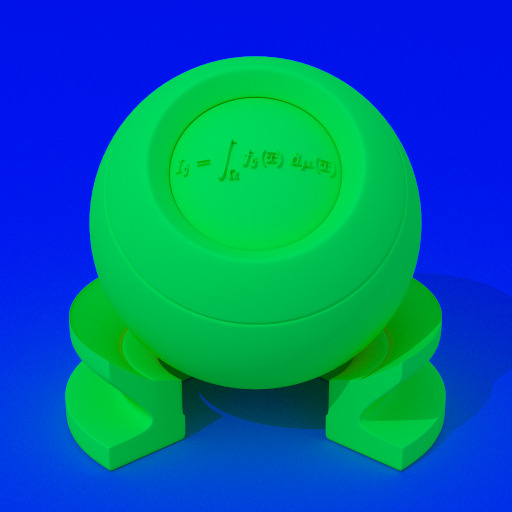}};
            \node[anchor=south west, inner sep=0] (herpipin_path_ours_3) at (0, 0                       ) {\includegraphics[height=\heightPath pt]{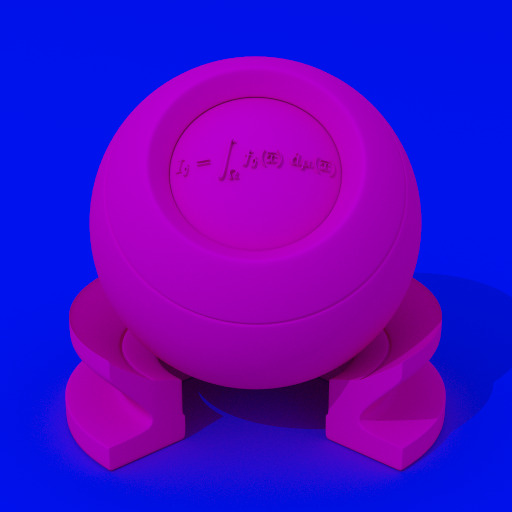}};
        \end{scope}

        \begin{scope}[xshift=\widthPathWithMargin pt]
            \clip (0 pt, 0 pt) rectangle +(.5*\widthPath pt, \teaserHeight pt);
            \node[anchor=south west, inner sep=0] (ixcclye_path_ref)  at (0, \heightPathWithMargin pt) {\includegraphics[height=\heightPath pt]{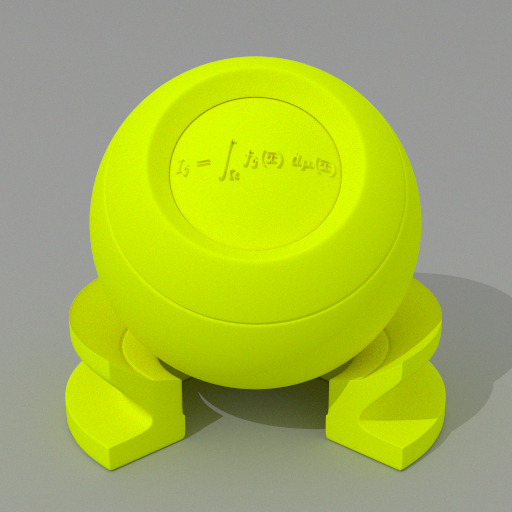}};
            \node[anchor=south west, inner sep=0] (herpipin_path_ref) at (0, 0                       ) {\includegraphics[height=\heightPath pt]{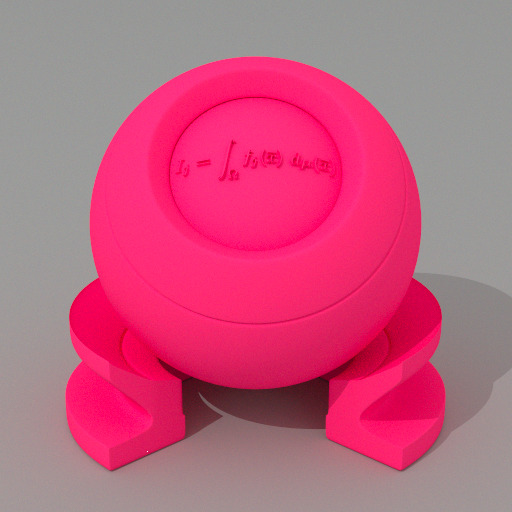}};
        \end{scope}

        \begin{scope}[xshift=\widthPathWithMargin pt]
            \clip[xshift=.5*\widthPath pt] (0 pt, 0 pt) rectangle +(.5*\widthPath pt, \teaserHeight pt);
            \node[anchor=south west, inner sep=0] (ixcclye_path_ref)  at (0, \heightPathWithMargin pt) {\includegraphics[height=\heightPath pt]{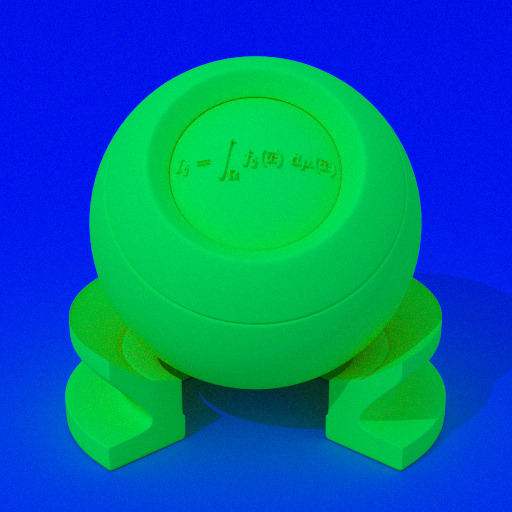}};
            \node[anchor=south west, inner sep=0] (herpipin_path_ref) at (0, 0                       ) {\includegraphics[height=\heightPath pt]{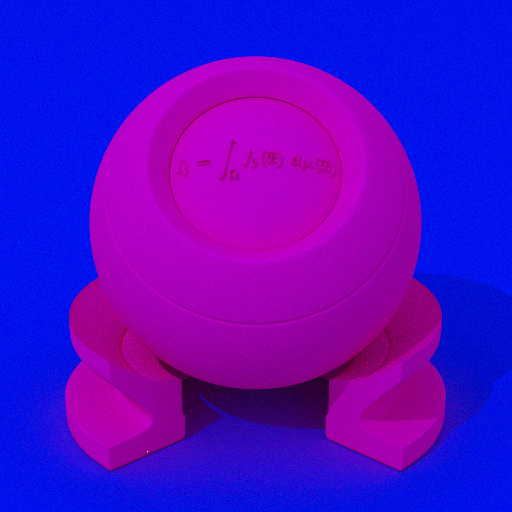}};
        \end{scope}

        \begin{scope}[xshift=2*\widthPathWithMargin pt]
            \clip (0 pt, 0 pt) rectangle +(.5*\widthPath pt, \teaserHeight pt);
            \node[anchor=south west, inner sep=0] (ixcclye_path_hullin_3)  at (0, \heightPathWithMargin pt) {\includegraphics[height=\heightPath pt]{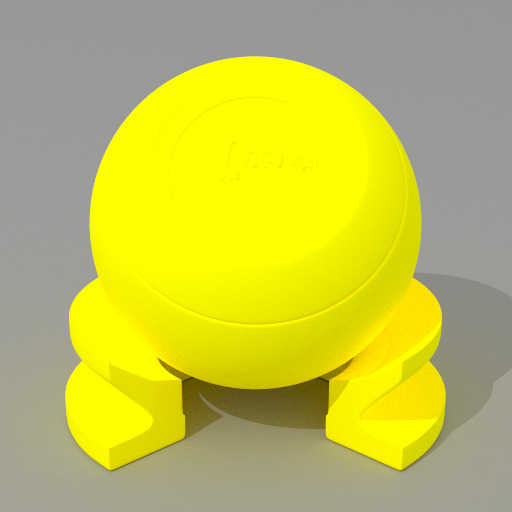}};
            \node[anchor=south west, inner sep=0] (herpipin_path_hullin_3) at (0, 0                       ) {\includegraphics[height=\heightPath pt]{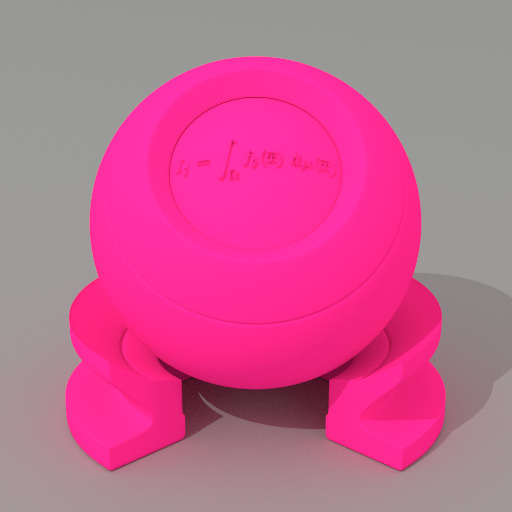}};
        \end{scope}

        \begin{scope}[xshift=2*\widthPathWithMargin pt]
            \clip[xshift=.5*\widthPath pt] (0 pt, 0 pt) rectangle +(.5*\widthPath pt, \teaserHeight pt);
            \node[anchor=south west, inner sep=0] (ixcclye_path_hullin_3)  at (0, \heightPathWithMargin pt) {\includegraphics[height=\heightPath pt]{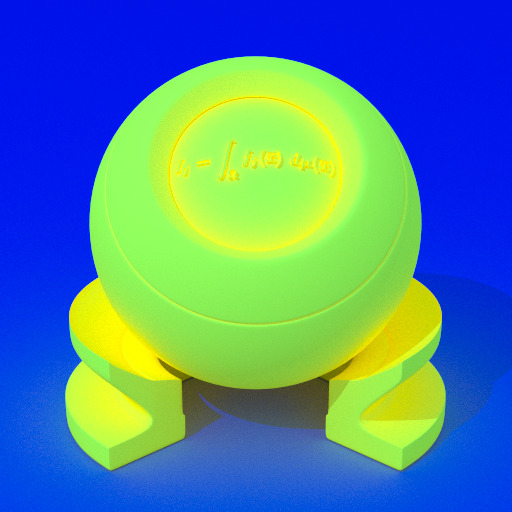}};
            \node[anchor=south west, inner sep=0] (herpipin_path_hullin_3) at (0, 0                       ) {\includegraphics[height=\heightPath pt]{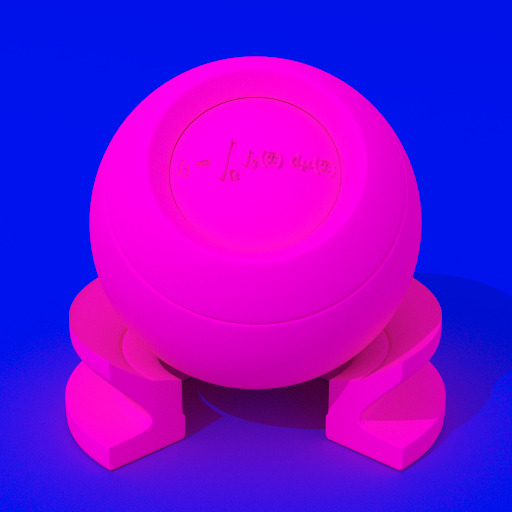}};
        \end{scope}

        \foreach \i in {0,...,2} {
            \tikzmath{
                \xshadow = \i * \widthPathWithMargin + .4;
            }
            \node[xshift=\xshadow               , anchor=north west, yshift=-.4pt, text opacity=.8] at (ixcclye_path_ours_3.north west) {\textcolor{gray} {\tiny{Daylight}}};
            \node[xshift=\i*\widthPathWithMargin, anchor=north west              , text opacity=.8] at (ixcclye_path_ours_3.north west) {\textcolor{white}{\tiny{Daylight}}};

            \node[xshift=\xshadow               , anchor=north east, yshift=-.4pt, text opacity=.8] at (ixcclye_path_ours_3.north east) {\textcolor{gray} {\tiny{Blue light}}};
            \node[xshift=\i*\widthPathWithMargin, anchor=north east              , text opacity=.8] at (ixcclye_path_ours_3.north east) {\textcolor{white}{\tiny{Blue light}}};

            \draw[xshift=\i*\widthPathWithMargin pt, line width=.1pt]                                  (0.5*\widthPath pt, 0 pt) -- (0.5*\widthPath pt, \heightPath pt);
            \draw[xshift=\i*\widthPathWithMargin pt, line width=.1pt, yshift=\heightPathWithMargin pt] (0.5*\widthPath pt, 0 pt) -- (0.5*\widthPath pt, \heightPath pt);

            \draw[xshift=\i*\widthPathWithMargin pt] (0, 0                       ) rectangle +(\widthPath pt, \heightPath pt);
            \draw[xshift=\i*\widthPathWithMargin pt] (0, \heightPathWithMargin pt) rectangle +(\widthPath pt, \heightPath pt);
        }

        \begin{pgfonlayer}{foreground}
            \node[anchor=base, yshift=5pt               ] (txtA) at (herpipin_path_ours_3.south)   {\textcolor{white}{Ours}};
            \node[anchor=base, yshift=5pt               ] (txtB) at (herpipin_path_ref.south)      {\textcolor{white}{Ref.}};
            \node[anchor=base, yshift=5pt               ] (txtC) at (herpipin_path_hullin_3.south) {\textcolor{white}{\edit{Naive}}};
            \node[anchor=base, yshift=-9pt               ] at (herpipin_path_ref.south)      {{(b) Offline rendering}};
        \end{pgfonlayer}
        \fill [fill=gray, fill opacity=0.6, rounded corners=5pt] (txtA.south west) + (0,1pt) rectangle (txtA.north east);
        \fill [fill=gray, fill opacity=0.6, rounded corners=5pt] (txtB.south west) + (0,1pt) rectangle (txtB.north east);
        \fill [fill=gray, fill opacity=0.6, rounded corners=5pt] (txtC.south west) + (0,1pt) rectangle (txtC.north east);
    \end{scope}

    \begin{scope}[xshift=\xStartLabels pt]
        \node[left]                                  at (\pathLabelsWidth pt, .5*\widthPath pt) {\rotatebox{90}{\scriptsize{\textsc{HERPIPIN}}}};
        \node[left, yshift=\heightPathWithMargin pt] at (\pathLabelsWidth pt, .5*\widthPath pt) {\rotatebox{90}{\scriptsize{\textsc{IXCCLYE}}}};
    \end{scope}
\end{tikzpicture}
    \vspace{-19pt}
    \centering
    \caption{
        \textbf{Fluorescent effects in a tristimulus render.}
    (a) We introduce a method to render fluorescent materials in a non-spectral rendererer, exhibiting typical {increased luminance and} hue-shifting effects when the same materials are placed under a different lighting.
    (b) Our method (left insets) is based on a new reduction technique that accurately matches the spectral reference (center insets), significantly improving on previous work \protect\Cite{Hullin10} (right insets) where \edit{a naive} formulation of the fluorescence reradiation matrix reduction was used.
    }
    \label{fig:teaser}
}

\maketitle

\begin{abstract}
We propose a method to accurately handle fluorescence in a non-spectral (\eg, tristimulus) rendering engine, showcasing color-shifting and increased luminance effects.
Core to our method is a principled reduction technique that encodes the re-radiation into a low-dimensional matrix working in the space of the renderer's Color Matching Functions (CMFs).
Our process is independent of a specific CMF set and allows for the addition of a non-visible ultraviolet band during light transport.
Our representation visually matches full spectral light transport for measured fluorescent materials even for challenging illuminants.


\end{abstract}


\section{Introduction}
Fluorescent materials absorb light and re-emit it at longer wavelengths. 
This effect has two significant implications in Computer Graphics.
First, fluorescence introduces color shifts: after an interaction with a fluorescent material, the light hue changes.
Second, the absorption/re-emission process may transfer light from the ultraviolet (UV) range toward the visual range.
This has the effect of increasing the perceived luminance of fluorescent materials, compared to other surrounding materials that only reflect light in the visual range.
Fluorescent effects are especially noticeable during dawn or dusk, which is due to Rayleigh scattering: sun light (depleted in UV) is less pronounced or absent, making light scattered from the sky (rich in UV) relatively more prominent.

Fluorescence thus has a significant potential to increase artistic control over a scene appearance~\cite{Jung19}.
However, the few rendering engines that \emph{accurately} support fluorescence (\eg, ART~\cite{ART}, Ocean~\cite{Ocean}) require spectral light transport, and assume this is a prerequisite for reproducing wavelength-shifting effects.
In contrast, traditional tristimulus light transport employs component-wise vector multiplications that do not readily model hue-shifting.
Fluorescence is instead merely mimicked in such rendering engines, most often through the addition of an emission term or the use of a higher albedo.
Not only may this break energy conservation, but it also neglects the impact of reflections from the surrounding environment on fluorescent appearance, and it entirely neglects light outside the visible spectral range.
Even though artists may manually adjust material parameters to visually reproduce a fluorescent appearance, this would be restricted to a specific lighting, forbidding light interaction such as shown in Figure~\ref{fig:teaser}~(a).

\if\finalversion1
\def\thefootnote{\star}\footnotetext{\vspace{-30pt}These authors contributed equally to this work}\def\thefootnote{\arabic{footnote}}
\fi

Our goal in this paper is to add support for the accurate reproduction of fluorescence in non-spectral (\eg, tristimulus) renderers, with a focus on Lambertian materials.
The central idea consists in replacing the classic vector-vector multiplication between the incoming radiance and diffuse albedo, with a matrix-vector multiplication where the matrix is obtained by a reduction of the fluorescence reradiation matrix.
A reduction approach has been suggested by Hullin~\etal~\cite{Hullin10}, but it is based on an empirical argument and produces inaccurate results (see Figure~\ref{fig:teaser}~(b), right insets). \edit{This} method injects unwanted energy since color matching functions (\eg, RGB or XYZ) are non-orthogonal.
Our main contribution lies in a principled reduction method, which relies on spectral upsampling and downsampling operators according to an arbitrary set of non-orthogonal functions and produces significantly more accurate results (see Figure~\ref{fig:teaser}~(b), left insets).
Since our method is generalizable to any basis, we show that an additional UV band is easily incorporated in our framework to account for reradiation from non-visible to visible light.

\section{Related work}
\paragraph*{Fluorescence rendering.}
With fluorescence, the reflected radiance of a material is no longer a mere function of the input wavelength $\li$, but also accounts for a reradiated wavelength $\lo$~\cite{Glassner95}:
\begin{align}
    L_o(\wo, \lo) = \iint \rho(\wi, \wo, \li, \lo) L_i(\wi, \li) \langle\wi,\wn\rangle\; \dd{\wi} \, \dd{\li},
    \label{eq:bispectral-reflection}
\end{align}
where $\rho(\wi, \wo, \li, \lo)$ is the Bidirectional Reflectance and Reradiation Distribution Functions (BRRDF), $L_o$ (resp. $L_i$) the outgoing (resp. incoming) radiance with $\wo$ (resp. $\wi$) the outgoing (resp. incoming) direction, and $\wn$ the shading normal.
Due to this formulation, fluorescence is restricted to spectral renderers, such as uni-directional path tracers~\cite{Wilkie01,Bending08,Mojzik18}, and
bi-directional path tracers thanks to wavelength mollification~\cite{Jung20}.
To the best of our knowledge, reproducing fluorescence in non-spectral renderers hasn't received much attention.

\paragraph*{Fluorescent models.}
Most BRRDF models assume a Lambertian reflectance~\cite{Jung18}, which confines fluorescence to a diffuse appearance.
A common way to represent fluorescence is through the reradiation (or Donaldson) matrix~\cite{Donaldson1954}.
We denote it by $\rerad(\li, \lo)$, with its horizontal and vertical axes corresponding respectively to the incoming and outgoing wavelengths.
The reflected radiance (Equation~\ref{eq:bispectral-reflection}) is now expressed as:
\begin{align}
    L_o(\wo, \lo) =  \int \frac{\rerad(\li, \lo)}{\pi} \int L_i(\wi, \li) \langle\wi,\wn\rangle\; \dd{\wi} \, \dd{\li}.
    \label{eq:bispectral-diffuse}
\end{align}
For a non-fluorescent material, $\rerad$ is a diagonal matrix holding the spectral diffuse albedo $\rho_d$ (\ie, $\rerad(\li, \lo) = \delta(\li-\lo) \rho_d(\li)$).

Hullin~\etal~\cite{Hullin10} modeled the BRRDF by an expansion into a series of products between angular and spectral functions, using a Principal Component Analysis (PCA).
This goes beyond diffuse fluorescence, and allows for the rendering of data-driven fluorescent glossy materials.
Recently, Benamira and Pattanaik~\cite{Benamira23} have extended the Microfacet theory~\cite{Torrance67} with an analytic fluorescent reflectance term.
While our formulation should extend to non-diffuse fluorescence, we leave it to future work and focus on Lambertian appearance.

Our approach is based on spectral upsampling and downsampling operators. However, this should not be confused with the spectral upsampling method of Jung~\etal~\cite{Jung19}: in their approach, upsampling is used to create a new fluorescent material that yields vivid colors, while in ours it is used to accurately account for fluorescent effects of an \emph{existing} material.

\paragraph*{Fluorescence acquisition.}
One way to measure the reradiation matrix is to illuminate a sample with a monochromatic wavelength while a spectral sensor captures the spectral reflectance response.
Such devices require costly and well-calibrated equipment, and are practically restricted to a single incoming-outgoing direction pair.
Gonzalez and Fairchild~\cite{Gonzalez00} measured a large range of Lambertian samples using such a device (Labsphere BFC-450); we evaluate our method based on their database.
The more recent work of Iser~\etal~\cite{Iser23} (which improves on the method of Blasinski~\etal~\cite{Blasinski20}) simplify the measurement process by using a Gaussian Mixture Model in place of a dense reradiation model. Our method could be equally applied to their database.

The method of Hullin~\etal~\cite{Hullin10} tackles the challenge of measuring both angular and spectral variations in a BRRDF.
To this end, they simplify the acquisition process by measuring dense spectral data for a few angular configurations, and dense angular data with a reduced spectral sampling.
They then combine these measurements through the aforementioned dedicated BRRDF model based on a series expansion.

\section{Naive Reradiation Matrix Reduction}
\label{sec:naive}

%
In an RGB renderer, we track a triplet of outgoing and incoming radiance $\mathbf{c}_o$ and $\mathbf{c}_i$ in place of the spectral quantity $L_o$ and $L_i$:
%
\begin{align}
    \label{eq:co}
    \mathbf{c}_o(\wo) = \int L_o(\wo, \lo) \, \mathbf{s}(\lo) \dd{\lo},
\end{align}
where $\mathbf{s}$ is a vector of sensitivity functions, one per color channel.
The definition for $\mathbf{c}_i$ is similar.
In the case of a non-fluorescent Lambertian material with an albedo vector $\pmb{\rho}_d$, the rendering equation is approximated using the component-wise product $\odot$:
\begin{align}
    \mathbf{c}_o(\wo) \approx \frac{1}{\pi} \int \pmb{\rho}_d \odot \mathbf{c}_i(\wi) \langle\wi,\wn\rangle\; \dd{\wi}.
\end{align}
%
By analogy, \edit{one} can approximate the fluorescent counterpart (Equation~\ref{eq:bispectral-diffuse}) using vector-matrix multiplications:
\edit{
\begin{align}
    \label{eq:solution-hullin}
    \mathbf{c}_o(\wo) \approx \frac{1}{\pi} \int \reduxh \times \mathbf{c}_i(\wi) \langle\wi,\wn\rangle\; \dd{\wi},
\end{align}
}
where \edit{$\reduxh$} is a reduced $3 \times 3$ reradiation matrix that accounts for reradiation from one color channel to the other. \edit{To render their BRRDF measurements, Hullin et al.~\cite{Hullin10} suggested to} construct such a reduced matrix by integrating the reradiation:
\begin{align}
    \reduxh_{k,l} = \iint \bar{s}_k(\li) \bar{s}_l(\lo) \rerad(\li, \lo) \dd{\li}\dd{\lo},
\end{align}
where the $\bar{s}_{k}(\lambda) = {s_{k}(\lambda) \over || s_k ||}$, $k \in \{r,g,b\}$ are normalized RGB Color Matching Functions (CMFs).
In matrix form, this yields:
\begin{align}
    \reduxh = \bar{S}^\top \times \rerad \times \bar{S}
    \label{eq:redux_hullin}
\end{align}
where $\bar{S}=[\bar{s}_r,\bar{s}_g,\bar{s}_b]$ is the $N \times 3$ matrix of normalized RGB sensitivity functions (with $N=431$ wavelength bins in their case).

\begin{figure}[t]
    \begin{tikzpicture}[font=\footnotesize]
        \tikzmath{
            \margins = 6;
            \widthImage = (\the\linewidth - \margins) / 2;
        }

        \node[inner sep=0, anchor=west] (matrix_3_hullin) at (0, 0) {\includegraphics[width=\widthImage pt]{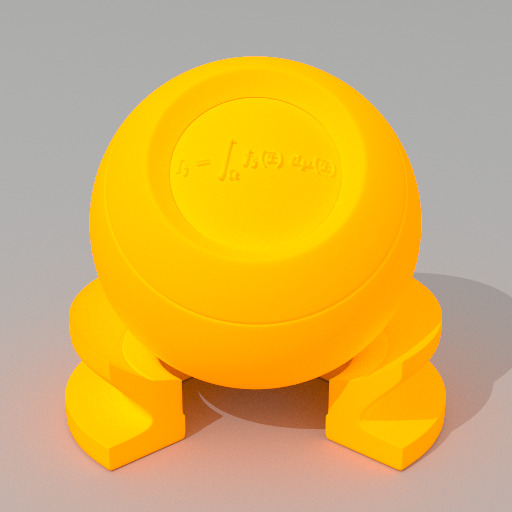}};
        \draw (matrix_3_hullin.south west) rectangle (matrix_3_hullin.north east);
        \node[below=3pt] at (matrix_3_hullin.south) {\textbf{(a) Naive Reduction}};

        \node[inner sep=0, anchor=east] (spectral) at (\linewidth, 0) {\includegraphics[width=\widthImage pt]{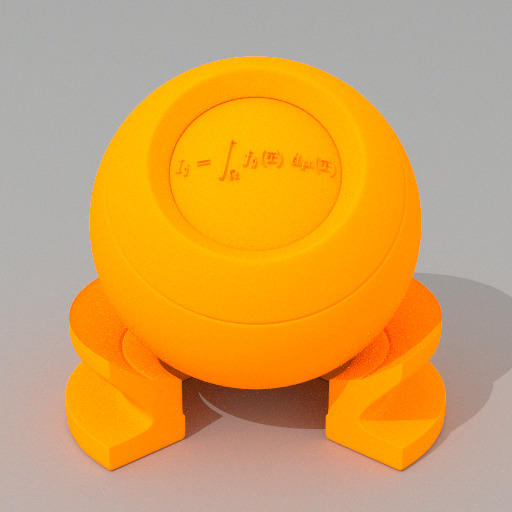}};
        \draw (spectral.south west) rectangle (spectral.north east);
        \node[below=3pt] at (spectral.south) {\textbf{(b) Spectral Reference}};
    \end{tikzpicture}
    \vspace{-20pt}
    \caption{
        \textbf{Naive reduction.} Using (a) the reduction method of \edit{Equation~\ref{eq:redux_hullin}} does not correctly capture the color of the fluorescent material compared to (b) the spectral reference. We use the \textnormal{\textsc{HERPIOYE}} material lit by a \textnormal{\textsc{D65}} spectrum.
        \label{fig:limits_naive_reduction}
        \vspace{-20pt}
    }
\end{figure}

\paragraph*{Limits of Naive Reduction.}
As shown by Hullin~\etal, this naive reduction technique does not faithfully reproduce fluorescent materials (see Figure~\ref{fig:limits_naive_reduction}), from which they conclude that dense Donaldson matrices and spectral rendering are required.

We do not share their conclusion and show that \edit{a different formulation can more correctly reproduce fluorescence.}
Indeed, \edit{with this naive approach}, an identity Donaldson matrix does not yield an identity reduced matrix, and introduces unwanted energy (see the first row of Figure~\ref{fig:results_patches}).
In the next section, we show that the correct construction of the reduced matrix is devoid of this issue, and hence produces much more faithful renderings.

\section{Correct Reradiation Matrix Reduction}

Here, we show that the correct way to model the reduced matrix is through spectral upsampling \& downsampling with non-orthogonal bases.
We start by describing our approach using XYZ triplets, as the corresponding sensitivity functions are non-negative.

\subsection{Reduction through upsampling \& downsampling}
\begin{figure*}[t!]
\hspace{-0.5cm}

\pgfdeclarelayer{background}
\pgfdeclarelayer{foreground}
\pgfsetlayers{background,main,foreground}

\resizebox{\linewidth}{!}{
\begin{tikzpicture}[font=\footnotesize]
    \tikzmath{
        \widthPlots=57;
        \heightPlots=\widthPlots / 2;
    }



    \pgfplotstableread[col sep=comma]{figures/test/reduced/assets/illu_spec.csv}\csvIlluRef
    \pgfplotstableread[col sep=comma]{figures/test/reduced/assets/illu_up.csv}\csvIlluUp
    \pgfplotstableread[col sep=comma]{figures/test/reduced/assets/c_o_up.csv}\csvReflUp
    \pgfplotstableread[col sep=comma]{figures/test/reduced/assets/c_o_spec.csv}\csvReflSpec
    \pgfplotstableread[col sep=comma]{figures/scene/cmf/ciexyz06_2deg.csv}\csvCMFs

    \node (Li) at (1.0cm,-10pt) {$L_i(\lambda)$};

    \begin{axis}[
        at=(Li.north),
        name=LiPlot,
        anchor=south,
        enlargelimits=false,
        width=\widthPlots pt,
        height=\heightPlots pt,
        scale only axis,
        grid=both,
        xmin=300,
        xmax=780,
        ymin=-0.01,
        ymax=150,
        xticklabel=\empty,
        yticklabel=\empty]
        \addplot[thick,color=tab_orange,forget plot,] table[x index = {0}, y index = {1}]{\csvIlluRef};
    \end{axis}


    \draw[->,color=tab_orange] (LiPlot.west) -- +(-1cm, 0) node [black, anchor=east] (dLi) {$\mathrm{d}\lambda_i$};

    \begin{axis}[
        at=(dLi.west),
        name=LiUpPlot,
        anchor=east,
        enlargelimits=false,
        width=\widthPlots pt,
        height=\heightPlots pt,
        scale only axis,
        axis background/.style={fill=white},
        grid=both,
        xmin=300,
        xmax=780,
        ymin=-0.01,
        ymax=150,
        xticklabel=\empty,
        yticklabel=\empty]
        \addplot[color=tab_orange,forget plot] table[x index = {0}, y index = {1}]{\csvIlluRef};
        \addplot[thick,color=tab_purple,forget plot] table[x index = {0}, y index = {1}]{\csvIlluUp};
    \end{axis}

    \node[anchor=east] (LiTimes) at (LiUpPlot.west) {$\times$};

    \node[anchor=east, inner sep=0] (rerad) at (LiTimes.west) {\includegraphics[height=\heightPlots pt]{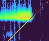}};
    \draw (rerad.south west) rectangle (rerad.north east);
    \node[anchor=north] (reradLabel) at (rerad.south) {$\rerad$};

    \node[anchor=east] (bigint) at (rerad.west) {$\bigint$};
    \node[anchor=east, xshift=-6pt] (equalLi) at (bigint) {$=$};

    \begin{axis}[
        at=(equalLi.west),
        xshift=-6pt,
        name=reflPlot,
        anchor=east,
        enlargelimits=false,
        width=\widthPlots pt,
        height=\heightPlots pt,
        scale only axis,
        axis background/.style={fill=white},
        grid=both,
        xmin=300,
        xmax=780,
        ymin=-0.01,
        ymax=180,
        xticklabel=\empty,
        yticklabel=\empty]
        \addplot[thick ,color=tab_orange,forget plot] table[x index = {0}, y index = {1}]{\csvReflSpec};
        \addplot[color=tab_purple,forget plot] table[x index = {0}, y index = {1}]{\csvReflUp};
    \end{axis}

    \begin{pgfonlayer}{background}
        \draw [fill=tab_blue, fill opacity=0.15, color=tab_blue]
            let \p0=(reflPlot.west)    in (\x0, \y0) 
            let \p1=(reradLabel.south) in (\x1, \y1) 
            let \p2=(dLi.east)         in (\x2, \y2) 
            let \p3=(LiUpPlot.north)   in (\x3, \y3) 
            ($(\x0, \y1) + (-6pt, 0pt)$)
            rectangle
            ($(\x2, \y3) + (6pt, 6pt)$);

        \draw [color=tab_blue, thick, decoration={brace,mirror,raise=0.25cm}, decorate, opacity=0.5]
            let \p0=(reflPlot.west)    in (\x0, \y0) 
            let \p1=(reradLabel.south) in (\x1, \y1) 
            let \p2=(dLi.east)         in (\x2, \y2) 
            let \p3=(LiUpPlot.north)   in (\x3, \y3) 
            ($(\x0, \y1) + (-6pt, 0pt)$) -- ($(\x2, \y1)+ (6pt, 0pt)$);
    \end{pgfonlayer}


    \draw[->, dashed, color=tab_orange] (reflPlot.west) -- +(-1cm, 0) node [black, anchor=east] (dLo) {$\mathrm{d}\lambda_o$};

    \begin{axis}[
        at=(dLo.west),
        name=reflPlotCmf,
        anchor=east,
        enlargelimits=false,
        width=\widthPlots pt,
        height=\heightPlots pt,
        scale only axis,
        axis background/.style={fill=white},
        grid=both,
        xmin=300,
        xmax=780,
        ymin=-0.01,
        ymax=180,
        xticklabel=\empty,
        yticklabel=\empty]
        \addplot[color=tab_orange ,forget plot] table[x index = {0}, y index = {1}]{\csvReflSpec};
        \addplot[thick,color=tab_purple,forget plot] table[x index = {0}, y index = {1}]{\csvReflUp};
    \end{axis}

    \node[anchor=east] (cmfTimes) at (reflPlotCmf.west) {$\times$};

    \begin{axis}[
        at=(cmfTimes.west),
        name=cmfPlot,
        anchor=east,
        enlargelimits=false,
        width=\widthPlots pt,
        height=\heightPlots pt,
        scale only axis,
        axis background/.style={fill=white},
        grid=both,
        xmin=300,
        xmax=780,
        ymin=-0.01,
        ymax=2,
        xticklabel=\empty,
        yticklabel=\empty]
        \addplot[thick,color=tab_red  ,forget plot] table[x index = {0}, y index = {1}]{\csvCMFs};
        \addplot[thick,color=tab_green,forget plot] table[x index = {0}, y index = {2}]{\csvCMFs};
        \addplot[thick,color=tab_blue ,forget plot] table[x index = {0}, y index = {3}]{\csvCMFs};
    \end{axis}
    \node[anchor=north] (cmfLabel) at (cmfPlot.south) {$\mathbf{s}(\lambda)$} ;

    \node[anchor=east] (bigint) at (cmfPlot.west) {$\bigint$};
    \node[anchor=east, xshift=-12pt] (equalCMF) at (bigint) {$=$};

    \begin{pgfonlayer}{background}
        \draw [fill=tab_green, fill opacity=0.15, color=tab_green]
            let \p0=(bigint.west)      in (\x0, \y0) 
            let \p1=(reradLabel.south) in (\x1, \y1) 
            let \p2=(dLo.east)         in (\x2, \y2) 
            let \p3=(LiUpPlot.north)   in (\x3, \y3) 
            ($(\x0, \y1) + (0pt, 0pt)$)
            rectangle
            ($(\x2, \y3) + (6pt, 6pt)$);

        \draw [color=tab_green, thick, decoration={brace,mirror,raise=0.25cm}, decorate, opacity=0.5]
            let \p0=(bigint.west)      in (\x0, \y0) 
            let \p1=(reradLabel.south) in (\x1, \y1) 
            let \p2=(dLo.east)         in (\x2, \y2) 
            let \p3=(LiUpPlot.north)   in (\x3, \y3) 
            ($(\x0, \y1) + (0pt, 0pt)$) -- ($(\x2, \y1)+ (6pt, 0pt)$);
    \end{pgfonlayer}


    \tikzmath{
        \heightSq=\heightPlots/3;
    }

    \draw[fill=white!79!black, anchor=east]
        let \p1=(cmfPlot.north) in (0,\y1)
        let \p2=(equalCMF.west) in (\x2, 0)
        (\x2, \y1) rectangle +(-\heightSq pt, -\heightSq pt);
    \draw[fill=white!43!black, anchor=east]
        let \p1=(cmfPlot.north) in (0,\y1)
        let \p2=(equalCMF.west) in (\x2, 0)
        ($(\x2, \y1) + (0, -\heightSq pt)$) rectangle +(-\heightSq pt, -\heightSq pt);
    \draw[fill=white!16!black, anchor=east]
        let \p1=(cmfPlot.north) in (0,\y1)
        let \p2=(equalCMF.west) in (\x2, 0)
        ($(\x2, \y1) + (0, -2*\heightSq pt)$) rectangle +(-\heightSq pt, -\heightSq pt);

    \node[anchor=east, xshift=-\heightSq pt] (colSpec) at (equalCMF.west) {$\begin{bmatrix}x \\y \\z\end{bmatrix}$};


    \tikzmath{
        \yOffset=42;
    }

    \draw[fill=white! 88!black] ($(LiPlot.south) - (0.5 * \heightSq pt, \yOffset               pt)$) rectangle +(\heightSq pt, -\heightSq pt);
    \draw[fill=white! 92!black] ($(LiPlot.south) - (0.5 * \heightSq pt, \yOffset +   \heightSq pt)$) rectangle +(\heightSq pt, -\heightSq pt);
    \draw[fill=white!100!black] ($(LiPlot.south) - (0.5 * \heightSq pt, \yOffset + 2*\heightSq pt)$) rectangle +(\heightSq pt, -\heightSq pt);
    \node[baseline, anchor=base] at ($(LiPlot.south) - (0, \yOffset + 4*\heightSq pt)$) {$\mathbf{c}_i$};


    \draw[fill=white! 88!black] ($(LiUpPlot.south) - (0.5 * \heightSq pt, \yOffset               pt)$) rectangle +(\heightSq pt, -\heightSq pt);
    \draw[fill=white! 92!black] ($(LiUpPlot.south) - (0.5 * \heightSq pt, \yOffset +   \heightSq pt)$) rectangle +(\heightSq pt, -\heightSq pt);
    \draw[fill=white!100!black] ($(LiUpPlot.south) - (0.5 * \heightSq pt, \yOffset + 2*\heightSq pt)$) rectangle +(\heightSq pt, -\heightSq pt);
    \node[baseline, anchor=base] (ciRefl) at ($(LiUpPlot.south) - (0, \yOffset + 4*\heightSq pt)$) {$\mathbf{c}_i$};

    \node[inner sep=0] (reradRGB) at ($(rerad.south) - (0, \yOffset + 1.5 * \heightSq pt)$) {\includegraphics[height=\heightPlots pt]{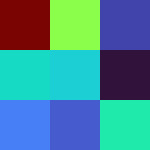}};
    \draw (reradRGB.south west) rectangle (reradRGB.north east);
    \node[baseline, anchor=base] at ($(rerad.south) - (0, \yOffset + 4*\heightSq pt)$) {$\reduxo$};

    \node[anchor=center] at ($(LiTimes.south) - (-\heightSq pt, \yOffset + 2.5 * \heightSq pt)$) {$\times$};
    \node[anchor=center] at ($(equalLi.south) - (0, \yOffset + 2.5 * \heightSq pt)$) {$=$};

    \draw[fill=white!65!black] ($(reflPlot.south) - (0.5 * \heightSq pt, \yOffset               pt)$) rectangle +(\heightSq pt, -\heightSq pt);
    \draw[fill=white!36!black] ($(reflPlot.south) - (0.5 * \heightSq pt, \yOffset +   \heightSq pt)$) rectangle +(\heightSq pt, -\heightSq pt);
    \draw[fill=white!16!black] ($(reflPlot.south) - (0.5 * \heightSq pt, \yOffset + 2*\heightSq pt)$) rectangle +(\heightSq pt, -\heightSq pt);
    \node[baseline, anchor=base] (c0Refl) at ($(reflPlot.south) - (0, \yOffset + 4*\heightSq pt)$) {$\mathbf{c}$};

    \begin{pgfonlayer}{background}
        \draw [fill=tab_blue, fill opacity=0.15, color=tab_blue]
            let \p0=(c0Refl.west)    in (\x0, \y0) 
            let \p1=(c0Refl.south)   in (\x1, \y1) 
            let \p2=(ciRefl.east)    in (\x2, \y2) 
            let \p3=(reradRGB.north) in (\x3, \y3) 
            ($(\x0, \y1) + (-6pt, 0pt)$)
            rectangle
            ($(\x2, \y3) + (6pt, 6pt)$);
    \end{pgfonlayer}

    \draw[->, color=tab_purple]
        let \p1=(LiPlot.south)   in (\x1, \y1)
        let \p2=(ciRefl.east)    in (\x2, \y2) 
        ($(LiPlot.south) - (0.75 * \heightSq pt, \yOffset + 1.5*\heightSq pt)$) -- (\x2, \y1 -\yOffset - 1.5*\heightSq pt);

    \draw[->, dashed, color=tab_purple]
    ($(reflPlot.south)    - ( 0.75*\heightSq pt, \yOffset + 1.5*\heightSq pt)$) --
    ($(reflPlotCmf.south) - (-0.75*\heightSq pt, \yOffset + 1.5*\heightSq pt)$);


    \draw[fill=white!65!black] ($(reflPlotCmf.south) - (0.5 * \heightSq pt, \yOffset               pt)$) rectangle +(\heightSq pt, -\heightSq pt);
    \draw[fill=white!36!black] ($(reflPlotCmf.south) - (0.5 * \heightSq pt, \yOffset +   \heightSq pt)$) rectangle +(\heightSq pt, -\heightSq pt);
    \draw[fill=white!16!black] ($(reflPlotCmf.south) - (0.5 * \heightSq pt, \yOffset + 2*\heightSq pt)$) rectangle +(\heightSq pt, -\heightSq pt);
    \node[baseline, anchor=base] (c0Transform) at ($(reflPlotCmf.south) - (0, \yOffset + 4*\heightSq pt)$) {$\mathbf{c}_o$};

    \node[anchor=center] at ($(cmfTimes.south) - (0, \yOffset + 2.5 * \heightSq pt)$) {$\times$};

    \node[baseline, anchor=center] (matTransform) at ($(cmfPlot.south) - (0, \yOffset + 1.5* \heightSq pt)$) {$T$};

    \begin{pgfonlayer}{background}
        \draw [fill=tab_green, fill opacity=0.15, color=tab_green]
            let \p0=(matTransform.west) in (\x0, \y0) 
            let \p1=(c0Refl.south)      in (\x1, \y1) 
            let \p2=(c0Transform.east)  in (\x2, \y2) 
            let \p3=(reradRGB.north)    in (\x3, \y3) 
            ($(\x0, \y1) + (-6pt, 0pt)$)
            rectangle
            ($(\x2, \y3) + (6pt, 6pt)$);
    \end{pgfonlayer}


    \node[anchor=east, xshift=-12pt] (equalCol) at ($(bigint) - (0, \yOffset + 3 * \heightSq pt)$) {$=$};

    \draw[fill=white!65!black, anchor=east]
        let \p1=(cmfPlot.south) in (0,\y1)
        let \p2=(equalCMF.west) in (\x2, 0)
        ($(\x2, \y1) - (0, \yOffset pt)$) rectangle +(-\heightSq pt, -\heightSq pt);
    \draw[fill=white!36!black, anchor=east]
        let \p1=(cmfPlot.south) in (0,\y1)
        let \p2=(equalCMF.west) in (\x2, 0)
        ($(\x2, \y1) - (0, \yOffset + \heightSq pt)$) rectangle +(-\heightSq pt, -\heightSq pt);
    \draw[fill=white!16!black, anchor=east]
        let \p1=(cmfPlot.south) in (0,\y1)
        let \p2=(equalCMF.west) in (\x2, 0)
        ($(\x2, \y1) - (0, \yOffset + 2*\heightSq pt)$) rectangle +(-\heightSq pt, -\heightSq pt);

    \node[baseline, anchor=east, xshift=-\heightSq pt] (xyzCol) at (equalCol.west) {$\begin{bmatrix}x \\y \\z\end{bmatrix}$};


    \draw[thick, dotted, gray!50!white]
        let \p1=(LiPlot.east) in (\x1, \y1)
        let \p2=(xyzCol.west) in (\x2, \y2)
        let \p3=(reradLabel.north)  in (\x3, \y3) 
        let \p4=(c0Refl.south)    in (\x4, \y4) 
        ($(\x1, \y3) - (0,  0.5 * \yOffset + .5*\heightSq pt)$) -- ($(\x2, \y3) - (0, 0.5 * \yOffset + .5*\heightSq pt)$);

    \draw[tab_orange, thick, ->] (Li.south) -- +($(0, -\yOffset pt) + (0, 1.5 * \heightSq pt)$) node[midway, anchor=west] {\rotatebox{90}{\footnotesize{\textbf{down}}}};
    \draw[tab_purple, thick, ->] ($(LiUpPlot.south) - (0, \yOffset pt)$) -- +(0, \yOffset - 6pt) node[midway, anchor=west] {\rotatebox{90}{\footnotesize{\textbf{up}}}};
    \draw[tab_orange, thick, <-] ($(reflPlot.south) - (0, \yOffset pt)$) -- +(0, \yOffset - 6pt) node[midway, anchor=west] {\rotatebox{90}{\footnotesize{\textbf{down}}}};
    \draw[tab_purple, thick, ->] ($(reflPlotCmf.south) - (0, \yOffset pt)$) -- +(0, \yOffset - 6pt) node[midway, anchor=west] {\rotatebox{90}{\footnotesize{\textbf{up}}}};

    \node[inner sep=0, baseline, color=gray] at ($(colSpec.west) + (-6pt, 0)$) {\rotatebox{90}{spectral}};
    \node[inner sep=0, baseline, color=gray] at ($(xyzCol.west) + (-6pt, 0)$)  {\rotatebox{90}{reduced}};
\end{tikzpicture}
\vspace{-10pt}
}
\caption{
    \textbf{Reduced Light Transport through Upsampling \& Downsampling.}
    {We illustrate the mapping between reduced transport (bottom row) and its spectral counterpart (top row).
    At each surface interaction (blue boxes), multiplication by the reduced matrix $\reduxo$ accounts for upsampling of the incoming reduced radiance, multiplication by $\rerad$, and downsampling of the outgoing spectral radiance.
    We compare the spectra obtained with this process (in purple) to those resulting from exact spectral light transport (in orange).
    After potentially multiple surface interactions (dashed arrows), spectral radiance must be integrated with the XYZ CMFs $\mathbf{s}(\lambda)$ in the spectral pipeline (green boxes).
    In the reduced pipeline, this is directly done through multiplication with a matrix $T$, which is the identity for XYZ reduced light transport.}
    \label{fig:scheme_upsampling_downsampling}
    \vspace{-20pt}
}
\end{figure*}

In the following, we approximate the bi-spectral rendering equation by a series of spectral downsampling and upsampling (see Figure~\ref{fig:scheme_upsampling_downsampling}). While our approach is applicable to the general form (Equation~\ref{eq:bispectral-reflection}), we demonstrate it on the diffuse case of Equation~\ref{eq:bispectral-diffuse} for simplicity.

We start by describing our spectral upsampling and downsampling operators, using the CMFs as spectral bases. The $k$-th color channel after spectral downsampling is then defined as:
\begin{align}
    \down[f]_k =  \int f(\lo) \; s_k(\lo) \dd{\lo}.
    \label{eq:downsampling}
\end{align}
\edit{
We can discretize downsampling with $S=[s_x, s_y, s_z]$ the $N \times 3$ matrix of sensitivity functions, and $\mathbf{f}$ the $N$-D vector of the spectrum:
\begin{align}
    \down[\mathbf{f}] =  S^T \times \mathbf{f},
    \label{eq:downsampling_vec}
\end{align}
}
\edit{
Since the $s_k$ ($k \in \{x,y,z\}$) form a non-orthogonal basis, we must use its dual basis $\tilde{s}_l$ ($l \in \{x,y,z\}$) to define upsampling from a color vector $\mathbf{c} = \left[ \dots,  c_i, \dots \right]$:
\begin{align}
    \up[\mathbf{c}](\li) = \sum c_l \; \tilde{s}_l(\li).
    \label{eq:upsampling}
\end{align}
The dual basis also has an associated $N \times 3$ matrix $\tilde{S} = ({S} {S}^{\top})^{-1} {S}$ (i.e., verifying $S^{\top}\tilde{S} = \mathrm{I}_3$, the $3 \times 3$ identity matrix\footnote{\vspace{-10pt}Since we have $S^T \times \tilde{S} = I_3 \iff ({S} {S}^{\top})\times \tilde{S} = S \iff \tilde{S} = ({S} {S}^{\top})^{-1} {S}$.}) to obtain the vector formulation for upsampling:
\begin{align}
    \up[\mathbf{c}] = \tilde{S} \times \mathbf{c}.
    \label{eq:upsampling_vec}
\end{align}
}
\indent In a tristimulus renderer, instead of transporting spectral radiance, we transport tristimulus radiance
$\mathbf{c} = \down\left[ L(\omega, \cdot) \right]$,
as illustrated in Figure~\ref{fig:scheme_upsampling_downsampling}.
Using a downsampled incoming radiance $\mathbf{c}_i = \down\left[ L_i(\wi, \cdot) \right]$ as input, we approximate Equation~\ref{eq:bispectral-diffuse}  by upsampling $\mathbf{c}_i$ and multiplying the resulting spectrum by the reradiation matrix:
\begin{align}
    L_o(\wo, \lo) \approx \int {\rerad(\li, \lo) \over \pi} \int   \, \up\left[ \mathbf{c}_i \right](\li) \, \langle\wi,\wn\rangle\, \dd{\wi} \, \dd{\li},
\end{align}
with the final pixel color given by $\mathbf{c}_o = \down\left[ L_o(\wo, \cdot) \right]$, which is equivalent to Equation~\ref{eq:co}.
Since all operators are linear, we re-write $\mathbf{c}_o$ in matrix form by swapping the order of integrals:
\begin{align}
    \label{eq:solution-ours}
    \mathbf{c}_o \approx {1 \over \pi} \int \left( \reduxo \times \mathbf{c}_i \right)\, \langle\wi,\wn\rangle\, \dd{\wi},
\end{align}
with the matrix-vector product given by:
\begin{align}
    \reduxo \times \mathbf{c}_i = \down \left[ \int \rerad(\li, \cdot) \; \up[\mathbf{c}_i](\li) \; \dd{\li} \right].
    \label{eq:least_square}
\end{align}
\edit{
\indent Our reduced matrix is obtained by using the discretized forms provided in Equations~\ref{eq:downsampling_vec} and~\ref{eq:upsampling_vec} into Equation~\ref{eq:least_square}, yielding:
}
\begin{align}
    \reduxo = S^{\top} \times \rerad \times \tilde{S},
    \label{eq:solution}
\end{align}

\begin{figure}[t]
    \begin{tikzpicture}[font=\footnotesize]
        \tikzmath{
            \margins = 6;
            \widthImage = (\the\linewidth - \margins) / 2;
        }
        \node[inner sep=0, anchor=west] (ours) at (0, 0) {\includegraphics[width=\widthImage pt]{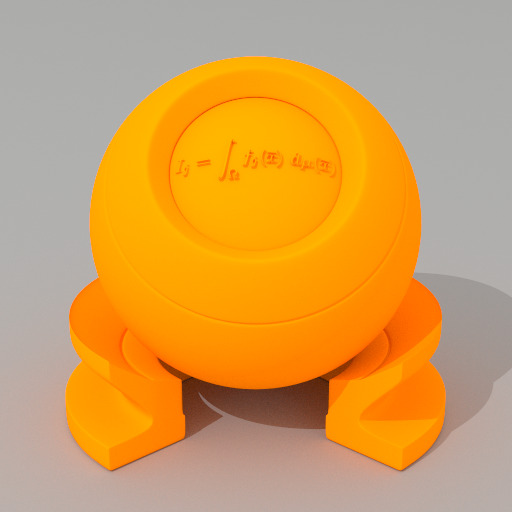}};
        \draw (ours.south west) rectangle (ours.north east);
        \node[below=3pt] at (ours.south) {\textbf{(a) Our Reduction}};

        \node[inner sep=0, anchor=east] (ref) at (\linewidth, 0) {\includegraphics[width=\widthImage pt]{figures/opt_reduced_matrix/HERPIOYE_D65_spectral_xyz.jpg}};
        \draw (ref.south west) rectangle (ref.north east);
        \node[below=3pt] at (ref.south) {\textbf{(b) Spectral Reference}};
    \end{tikzpicture}
    \vspace{-20pt}
    \caption{
        \textbf{Optimized reduction.} Using (a) our formulation, we obtain a new reduced re-radiation matrix that is better conditioned for rendering fluorescent materials compared to (b) the spectral reference, using the same material and illumination as in Figure~\ref{fig:limits_naive_reduction}.
        \label{fig:optimized_reradiation_matrix}
        \vspace{-20pt}
    }
\end{figure}

Equation~\ref{eq:solution-ours} has the same form as Equation~\ref{eq:solution-hullin}, but is derived from a principled approach that highlights the importance of using the dual basis.
Note that there is no need to normalize sensitivity functions with our approach.
More importantly, our solution in Equation~\ref{eq:solution} yields an identity reduced matrix when given an identity reradiation matrix.
Indeed, when $\rerad = \mathrm{I}_N$, $\reduxo = S^{\top} \tilde{S} = \mathrm{I}_3$.

We show in Figure~\ref{fig:optimized_reradiation_matrix} that our solution produces a closer match to the spectral reference.
In particular, it does not introduce unwanted energy.
As with \edit{the naive method}, our approach may be applied iteratively to build paths starting from the light source and ending at the camera.
However, our solution is justified thanks to our formulation based on downsampling and upsampling operators, as illustrated in Figure~\ref{fig:scheme_upsampling_downsampling}.
As shown in the top row, even though the upsampled spectra (in purple) do not initially match the exact spectra (in orange), they do closely reproduce them after a single multiplication by $\rerad$.
The main advantage of our method is that this accurate approximation is done implicitly through $\reduxo$ (blue boxes), as shown in the bottom row.
Note that $\reduxo$ will likely be entirely filled due to the wide basis support, even though $\rerad$ is close to upper triangular (although not exactly due to measurement errors).

The final step of the rendering process (green boxes in Figure~\ref{fig:scheme_upsampling_downsampling}) is to integrate spectra over XYZ CMFs.
For reduced light transport, this is done through a mere multiplication by a matrix $T$, which is the identity in the case of XYZ transport.
It might differ with a different choice of basis, as discussed in the next section where we consider the extension of the XYZ basis functions to account for the reradiation from the UV-range.

\subsection{Extending Reduction to Incorporate Non-Visible Spectra}

\begin{figure}[b]
    \vspace{-10pt}
    \setlength{\fboxsep}{0pt}
    \begin{tikzpicture}[font=\footnotesize]
        \tikzmath{
            \margins = 6;
            \widthImage = (\the\linewidth - 2*\margins) / 3;
            \xImage1 = 0;
            \xImage2 = \xImage1 + \widthImage + \margins;
            \xImage3 = \xImage2 + \widthImage + \margins;
            \heightPlot = .5 * \widthImage;
        }

        \node[inner sep=0, anchor=west] (matrix_3) at (\xImage1 pt, 0) {\includegraphics[width=\widthImage pt]{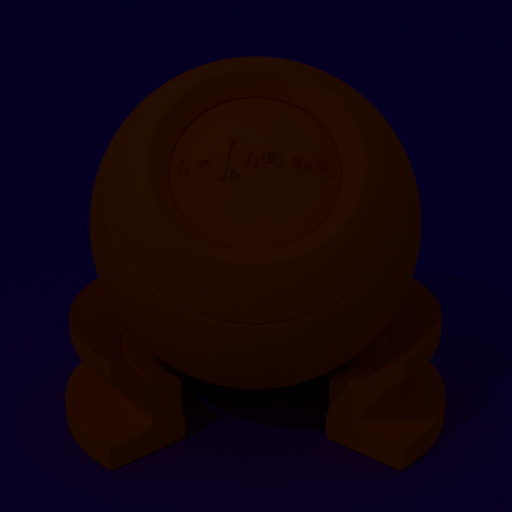}};
        \draw (matrix_3.south west) rectangle (matrix_3.north east);
        \pgfplotstableread[col sep=comma]{figures/scene/cmf/ciexyz06_2deg.csv}\mydata
        \begin{pgfinterruptboundingbox}
            \begin{axis}[
                at=(matrix_3.north),
                clip=true,
                anchor=south,
                yshift=\margins pt,
                width=\widthImage pt,
                height=\heightPlot pt,
                scale only axis,
                enlargelimits=false,
                grid=both,
                xmin=300,
                xmax=800,
                ymin=-0.01,
                ymax=2,
                xticklabel=\empty,
                yticklabel=\empty]
                \addplot[thick,color=tab_red  ,forget plot,] table[x index = {0}, y index = {1}]{\mydata};
                \addplot[thick,color=tab_green,forget plot,] table[x index = {0}, y index = {2}]{\mydata};
                \addplot[thick,color=tab_blue ,forget plot,] table[x index = {0}, y index = {3}]{\mydata};
            \end{axis}
        \end{pgfinterruptboundingbox}
        \node[below=3pt] at (matrix_3.south) {\textbf{(a) XYZ basis)}};

        \node[inner sep=0, anchor=west] (matrix_4) at (\xImage2 pt, 0) {\includegraphics[width=\widthImage pt]{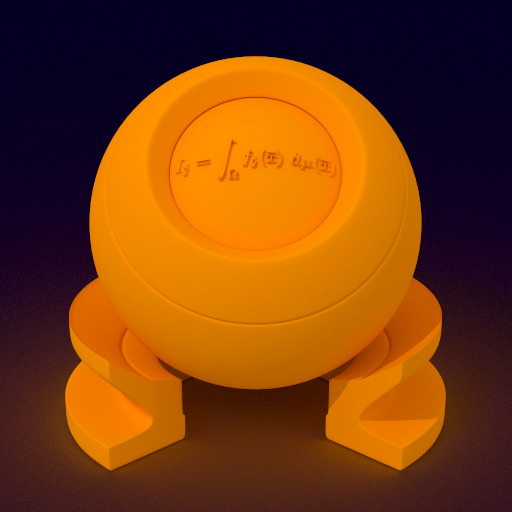}};
        \draw (matrix_4.south west) rectangle (matrix_4.north east);
        \pgfplotstableread[col sep=comma]{figures/scene/cmf/xyzu.csv}\mydata
            \begin{axis}[
                at=(matrix_4.north),
                clip=true,
                anchor=south,
                yshift=\margins pt,
                width=\widthImage pt,
                height=\heightPlot pt,
                scale only axis,
                enlargelimits=false,
                grid=both,
                xmin=300,
                xmax=800,
                ymin=-0.01,
                ymax=2,
                xticklabel=\empty,
                yticklabel=\empty]
                \addplot[thick,color=tab_red  ,forget plot,] table[x index = {0}, y index = {1}]{\mydata};
                \addplot[thick,color=tab_green,forget plot,] table[x index = {0}, y index = {2}]{\mydata};
                \addplot[thick,color=tab_blue ,forget plot,] table[x index = {0}, y index = {3}]{\mydata};
                \addplot[thick,color=tab_pink ,forget plot,] table[x index = {0}, y index = {4}]{\mydata};
            \end{axis}
        \node[below=3pt] at (matrix_4.south) {\textbf{(b) XYZU basis}};

        \node[inner sep=0, anchor=west] (reference) at (\xImage3 pt, 0) {\includegraphics[width=\widthImage pt]{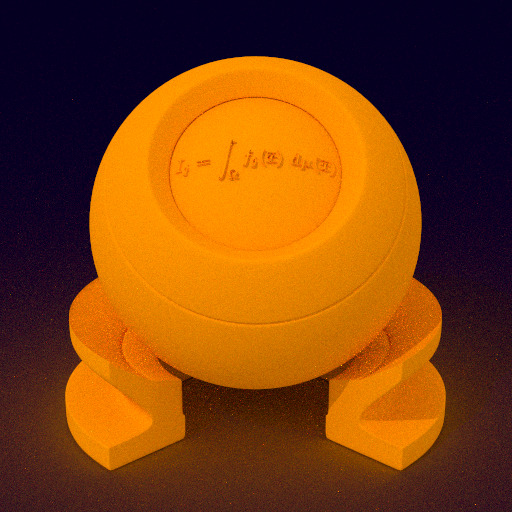}};
        \draw (reference.south west) rectangle (reference.north east);
        \node[below=3pt] at (reference.south) {\textbf{(c) Spectral Reference}};
    \end{tikzpicture}
    \vspace{-15pt}
    \caption{
        \textbf{Adding a UV channel.} Adding a fourth channel in the ultra-violet range, shown by the (b) pink curve, permits a more faithful reproduction of the (c) spectral reference when lit with UV lights. We used the \textnormal{\textsc{HERPIOYE}} material lit by a Gaussian spectrum centred at $350\:\mathrm{nm}$ with a standard deviation of $50\:\mathrm{nm}$.
        \label{fig:adding_uv_channels}
    }
\end{figure}

We can use Equations~\ref{eq:downsampling}~to~\ref{eq:solution} with an arbitrary set of non-orthogonal functions.
We experimented the use of a fourth sensitivity function to capture the UV-range.
To do so, we added the first element of a non-uniform B-spline Partition of Unity of $5$ elements and degree $2$ as detailed in Appendix~\ref{sec:alternate-bases} (the pink curve in Figure~\ref{fig:adding_uv_channels}~(b), top row).
During rendering, we need to transport an additional float with our XYZ triplet, which we denote by the term XYZU rendering.
Figure~\ref{fig:adding_uv_channels} shows the impact of this additional basis function to handle fluorescence from UV lighting.

However, when using an alternative basis, a final projection by a transfer matrix $T$ is required to project the n-uple in the rendering basis to the triplet of the CMF.
In the case of XYZU rendering:
$$
\begin{bmatrix}
    x \\ y \\ z
\end{bmatrix}
=
T \times \mathbf{c}_o,
\quad \mbox{with}\;
T =
\begin{bmatrix}
    1 & 0 & 0 & 0 \\
    0 & 1 & 0 & 0 \\
    0 & 0 & 1 & 0 \\
\end{bmatrix}.
$$
\vspace{-1pt}

\section{Implementation \& Results}
We validate our method in two rendering scenarios: in a GPU rasterizer for real-time performance; and inside a path tracer to reproduce a spectral reference and {test our method with} global illumination. Each implementation showcases a distinct methodology to compute light transport.

\subsection{Implementing Reduced Light Transport}
The reduced Rendering Equation (Equation~\ref{eq:solution-ours}) does not fully detail how we should solve light transport.
Depending on whether we solve light transport starting from the light or from the camera, either forward tracing or its adjoint must be implemented.
\paragraph*{Light Tracing.} This method should be used for photon mapping and direct lighting (see Section~\ref{sec:results_unity}). It is the simplest as it follows Equation~\ref{eq:solution-ours}: we downsample the light spectrum to
{a reduced radiance vector that is multiplied by a reduced reflectance matrix at each surface interaction, until it reaches the eye.}

\paragraph*{Adjoint Transport.} This method is required for backward path tracing (see Section~\ref{sec:results_path_tracing}).
\edit{
A difficulty here is that the adjoint of transporting reduced radiance vectors is to propagate a reduced matrix of throughputs (see Jarabo and Arellano~\cite{Jarabo2018} for details).
}
In this case, we start at the sensor with the identity matrix and multiply it at each surface interaction with the corresponding reduced reflectance {matrix}.
When next-event estimation queries the light,
{we simply multiply the downsampled light spectrum with the accumulated throughput matrix.}

\subsection{Results for Direct Lighting} \label{sec:results_unity}
\begin{figure*}[p!]
\center
\begin{tikzpicture}[font=\footnotesize]

   \tikzmath{
      \hmarginRightMatrix=16;
      \hmarginInnerIllu=6;
      \hmarginRightIllu=6;
      \vmargins=6;
      \innerSepSz = 3;
      \borderSz = 1.0;
      \lineHeight=56.0;
      \lineHeightWithMargin = \lineHeight + \vmargins;
      \startMatrix=12;
      \widthMatrix=49/41 * \lineHeight;
      \nIllu = 7;
      \startIllu = \startMatrix + \widthMatrix + \hmarginRightMatrix;
      \widthOneIllu = 0.5*\lineHeight;
      \widthHalfIllu = \widthOneIllu/2;
      \widthOneIlluWithMargin = \widthOneIllu + \hmarginInnerIllu;
      \widthIllu = (\nIllu - 1) * \widthOneIlluWithMargin + \widthOneIllu;
      \startMono = \startIllu + \widthIllu + \hmarginRightIllu;
      \swipeHeight = \lineHeight / 5;
      \widthMono = \the\linewidth - \startMono; 
   }

   \foreach \mat  [count = \xj] in {Identity, HERPICER, HERPIORA, IXCRLALE, TEXTYELL, POLGREE, PHP8HP1C, CIBA12, CIPLAW10} { 
      \begin{scope}[yshift=-\xj*\lineHeightWithMargin pt]

         \begin{scope}[xshift=\startMatrix pt]
            \node[inner sep=0, anchor=south west] (\mat_rerad) { \includegraphics[height=\lineHeight pt]{figures/patches/with_diag/\mat/rerad.jpg} };
            \draw (\mat_rerad.south west) rectangle (\mat_rerad.north east);
         \end{scope}

         \node[anchor=west] at (0, .5*\lineHeight pt) {\rotatebox{90}{\textsc{\mat}}};


         \begin{scope}[xshift=\startIllu pt]
            \foreach \illum  [count = \iidxIllum] in {A,E,D60,D65,FL1,FL2,HP5} {
               \tikzmath{
                     \idxIllum = \iidxIllum - 1;
               }
               \begin{scope}[xshift=\idxIllum * \widthOneIlluWithMargin pt]
                     \node[inner sep=0, anchor=south west] (\mat_ref_\illum) {\includegraphics[width=\widthOneIllu pt, height=\lineHeight pt]{figures/patches/with_diag/\mat/\illum_ref.jpg}};

                     \node[inner sep=0, anchor=south west] (\mat_ours_3_\illum)   at (.5*\widthOneIllu pt, .75*\lineHeight pt) {\includegraphics[width=\widthHalfIllu pt, height=\widthHalfIllu pt]{figures/patches/with_diag/\mat/\illum_ours_3.jpg}};
                     \node[inner sep=0, anchor=south west] (\mat_ours_4_\illum)   at (.5*\widthOneIllu pt, .50*\lineHeight pt) {\includegraphics[width=\widthHalfIllu pt, height=\widthHalfIllu pt]{figures/patches/with_diag/\mat/\illum_ours_4.jpg}};
                     \node[inner sep=0, anchor=south west] (\mat_hullin_4_\illum) at (.5*\widthOneIllu pt, .25*\lineHeight pt) {\includegraphics[width=\widthHalfIllu pt, height=\widthHalfIllu pt]{figures/patches/with_diag/\mat/\illum_hullin_4.jpg}};
                     \node[inner sep=0, anchor=south west] (\mat_hullin_3_\illum) at (.5*\widthOneIllu pt, 0.0pt)              {\includegraphics[width=\widthHalfIllu pt, height=\widthHalfIllu pt]{figures/patches/with_diag/\mat/\illum_hullin_3.jpg}};

                     \draw[line width=0.25pt] (\mat_ours_4_\illum.north east)   -- +(-\innerSepSz pt, 0);
                     \draw[line width=0.25pt] (\mat_hullin_4_\illum.north east) -- +(-\innerSepSz pt, 0);
                     \draw[line width=0.25pt] (\mat_hullin_3_\illum.north east) -- +(-\innerSepSz pt, 0);

                     \draw[line width=0.25pt] (\mat_ours_4_\illum.north west)   -- +(\innerSepSz pt, 0);
                     \draw[line width=0.25pt] (\mat_hullin_4_\illum.north west) -- +(\innerSepSz pt, 0);
                     \draw[line width=0.25pt] (\mat_hullin_3_\illum.north west) -- +(\innerSepSz pt, 0);

                     \draw[line width=0.25pt] (\mat_ours_4_\illum.north west)   -- +(0, \innerSepSz pt);
                     \draw[line width=0.25pt] (\mat_hullin_4_\illum.north west) -- +(0, \innerSepSz pt);
                     \draw[line width=0.25pt] (\mat_hullin_3_\illum.north west) -- +(0, \innerSepSz pt);
                     \draw[line width=0.25pt] (\mat_ours_4_\illum.north west)   -- +(0, -\innerSepSz pt);
                     \draw[line width=0.25pt] (\mat_hullin_4_\illum.north west) -- +(0, -\innerSepSz pt);
                     \draw[line width=0.25pt] (\mat_hullin_3_\illum.north west) -- +(0, -\innerSepSz pt);

                     \draw[line width=0.25pt] (\mat_ref_\illum.north) -- +(0, -\innerSepSz pt);
                     \draw[line width=0.25pt] (\mat_ref_\illum.south) -- +(0,  \innerSepSz pt);

                     \draw (\mat_ref_\illum.south west) rectangle (\mat_ref_\illum.north east);

               \end{scope}
            }
         \end{scope}


         \begin{scope}[xshift=\startMono pt]
            \node[inner sep=0, anchor=south west] (\mat_mono_hullin_3) at (0pt, 0.00*\lineHeight pt) {\includegraphics[width=\widthMono pt,height=\swipeHeight pt]{figures/patches/with_diag/\mat/mono_hullin_3.jpg}};
            \node[inner sep=0, anchor=south west] (\mat_mono_hullin_4) at (0pt, 0.20*\lineHeight pt) {\includegraphics[width=\widthMono pt,height=\swipeHeight pt]{figures/patches/with_diag/\mat/mono_hullin_4.jpg}};
            \node[inner sep=0, anchor=south west] (\mat_mono_ref)      at (0pt, 0.40*\lineHeight pt) {\includegraphics[width=\widthMono pt,height=\swipeHeight pt]{figures/patches/with_diag/\mat/mono_reference.jpg}};
            \node[inner sep=0, anchor=south west] (\mat_mono_ours_4)   at (0pt, 0.60*\lineHeight pt) {\includegraphics[width=\widthMono pt,height=\swipeHeight pt]{figures/patches/with_diag/\mat/mono_ours_4.jpg}};
            \node[inner sep=0, anchor=south west] (\mat_mono_ours_3)   at (0pt, 0.80*\lineHeight pt) {\includegraphics[width=\widthMono pt,height=\swipeHeight pt]{figures/patches/with_diag/\mat/mono_ours_3.jpg}};

            \draw[line width=0.25pt] (0pt, 0.20*\lineHeight pt) -- +(\widthMono pt, 0);
            \draw[line width=0.25pt] (0pt, 0.40*\lineHeight pt) -- +(\widthMono pt, 0);
            \draw[line width=0.25pt] (0pt, 0.60*\lineHeight pt) -- +(\widthMono pt, 0);
            \draw[line width=0.25pt] (0pt, 0.80*\lineHeight pt) -- +(\widthMono pt, 0);
            \draw                    (\mat_mono_hullin_3.south west) rectangle (\mat_mono_ours_3.north east);
        \end{scope}
      \end{scope}
   }


   \node[anchor=base, xshift=\startMatrix pt] at (.5*\widthMatrix pt, 16pt) {$\rerad$};
   \node[anchor=base, xshift=\startIllu pt]   at (.5*\widthIllu pt, 16pt) {\textbf{Lit by Standard Illuminants}};
   \node[anchor=base, xshift=\startMono pt]   at (.5*\widthMono pt, 16pt) {\textbf{Lit by Monochromatic Swipe}};

   \node[above=3pt, anchor=south, inner sep=0] (matrix_colormap) at (Identity_rerad.north) {\includegraphics[width=\widthMatrix pt]{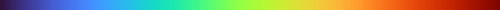}};
   \node[above] at (matrix_colormap.north) {\tiny{[dB]}};

   \draw (matrix_colormap.south west) rectangle (matrix_colormap.north east);
   \draw (matrix_colormap.north west) -- +(0, 3 pt) node[above] {\tiny{-30}};
   \draw (matrix_colormap.north east) -- +(0, 3 pt) node[above] {\tiny{0}};


   \node[xshift=-.25*\widthOneIllu pt] at (Identity_ref_A) {(r)};
   \node at (Identity_ours_3_A) {(a)};
   \node at (Identity_ours_4_A) {(b)};
   \node at (Identity_hullin_4_A) {(c)};
   \node at (Identity_hullin_3_A) {(d)};

   \foreach \illum  [count = \xi] in {A,E,D60,D65,FL1,FL2,HP5} {
      \begin{scope}[xshift=1.1*\xi pt]
         \node[above = 3pt] at (Identity_ref_\illum.north) {\textsc{\illum}};
      \end{scope}
   }

   \node[anchor=west] at (Identity_mono_ours_3.west)   {\textcolor{white}{\tiny{(a)}}};
   \node[anchor=west] at (Identity_mono_ours_4.west)   {\textcolor{white}{\tiny{(b)}}};
   \node[anchor=west] at (Identity_mono_ref.west)      {\textcolor{white}{\tiny{(r)}}};
   \node[anchor=west] at (Identity_mono_hullin_4.west) {\textcolor{white}{\tiny{(c)}}};
   \node[anchor=west] at (Identity_mono_hullin_3.west) {\textcolor{white}{\tiny{(d)}}};

   \node[above = 3pt, inner sep=0, anchor=south] (mono_illu) at (Identity_mono_ours_3.north) {\includegraphics[width=\widthMono pt,height=\swipeHeight pt]{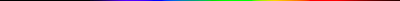}};
   \draw (mono_illu.south west) rectangle (mono_illu.north east);

   \draw (CIPLAW10_rerad.south west) -- +(0, -3pt) node[below] {\tiny{300}};
   \draw (CIPLAW10_rerad.south east) -- +(0, -3pt) node[below] {\tiny{780}};
   \node[below] at (CIPLAW10_rerad.south) {\tiny{$\lambda_i$ [nm]}};


   \draw (CIPLAW10_rerad.south east) -- +(3pt, 0) node[right] {\rotatebox{90}{\tiny{380}}};
   \draw (CIPLAW10_rerad.north east) -- +(3pt, 0) node[right] {\rotatebox{90}{\tiny{780}}};
   \node[right] at (CIPLAW10_rerad.east) {\rotatebox{90}{\tiny{$\lambda_o$ [nm]}}};

   \draw (CIPLAW10_mono_hullin_3.south west) -- +(0, -3pt) node[below] {\tiny{300}};
   \draw (CIPLAW10_mono_hullin_3.south east) -- +(0, -3pt) node[below] {\tiny{700}};
   \node[below] at (CIPLAW10_mono_hullin_3.south) {\tiny{$\lambda$ [nm]}};

\end{tikzpicture}
\caption{
   \textbf{Validation of our method with patches.}
   (r) Reference, (a) Ours XYZ, (b)  Ours XYZU, (c) \edit{Naive} XYZU, (d) \edit{Naive} XYZ.
   We display the reradiation matrix (left), color after one bounce for various illuminants (center) and a swipe with a delta illuminant (right).
   \label{fig:results_patches}
}
\end{figure*}

\edit{
\paragraph*{Unit Tests.} We compared our method with a spectral reference first by rendering patches of measured materials illuminated by different lights sources (see Figure~\ref{fig:results_patches}). On average, our method provides a smaller numerical difference than the naive method (in $\Delta E^*_{2000}$, see Table~\ref{tbl:deltaE}). See our supplemental for a detailed evaluation of every sample of the database of Gonzalez and Fairchild~\cite{Gonzalez00}.}
\begin{table}[t]
    \scriptsize
    \edit{
    \begin{tabularx}{\linewidth}{Xccccccc}
        \toprule
        &A&E&D60&D65&FL1&FL2&HP5\\
        \midrule
        XYZ Naive & 10.11& 9.73& 12.77& 12.88& 14.92& 10.88& 10.57\\
        XYZ Ours  & 3.92& 5.02& 3.88& 3.96& 1.14& 0.62& 2.48\\
        \midrule
        XYZU Naive& 10.56& 11.20& 13.43& 13.40& 15.18& 11.51& 11.22\\
        \textbf{XYZU Ours} & \textbf{3.86} & \textbf{3.23}& \textbf{3.36}& \textbf{3.33}& \textbf{1.04}& \textbf{0.52} & \textbf{2.43}\\
        \bottomrule
    \end{tabularx}
    \caption{\textbf{Average $\Delta E^*_{2000}$.} We excluded IXCAXORA from the database~\cite{Gonzalez00} due to inconsistencies in the measurement. \label{tbl:deltaE}
    \vspace{-12pt}
    }
    }
\end{table}

\paragraph*{Real-Time Results in Unity.} We further integrated our reduced fluorescent method in Unity using the High Definition Render Pipeline (HDRP) by modifying the Lit shader.
The working color space in Unity is RGB. We thus choose to convert the reduced reradiation from XYZ to RGB using:
\begin{align}
    \reduxo_{RGB} =  M_{XYZ\rightarrow RGB} \times \reduxo_{XYZ} \times M_{XYZ\rightarrow RGB}^{-1},
\end{align}
where $M_{XYZ\rightarrow RGB}$ converts a XYZ triplet to RGB.
We used the light tracing formulation and stored two additional \texttt{vec3} uniforms for the $6$ non-diagonal matrix elements. The only other modification is to change the component-wise vector multiplication between the albedo and the illuminant with a matrix-vector multiplication.

\begin{figure}[b!]
    \center
    \begin{tikzpicture}[font=\footnotesize]
        \tikzmath{
            \margins = 6;
            \widthImage = (\the\linewidth - \margins) / 2;
        }

        \node[inner sep=0, anchor=west] (matrix_path) at (0, 0) {\includegraphics[width=\widthImage pt]{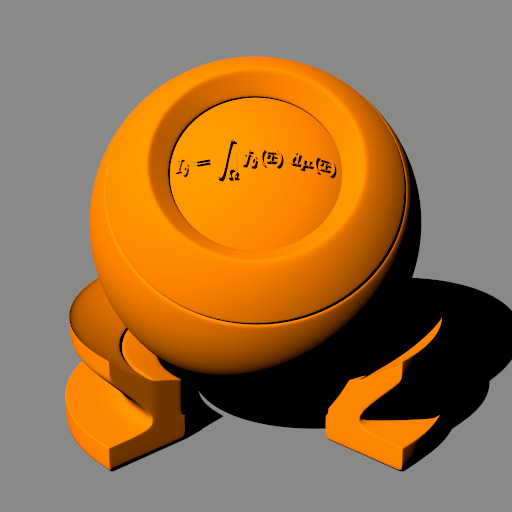}};
        \draw (matrix_3_hullin.south west) rectangle (matrix_3_hullin.north east);
        \node[below=3pt] at (matrix_3_hullin.south) {\textbf{(a) Path traced}};

        \node[inner sep=0, anchor=east] (matrix_unity) at (\linewidth, 0) {\includegraphics[width=\widthImage pt]{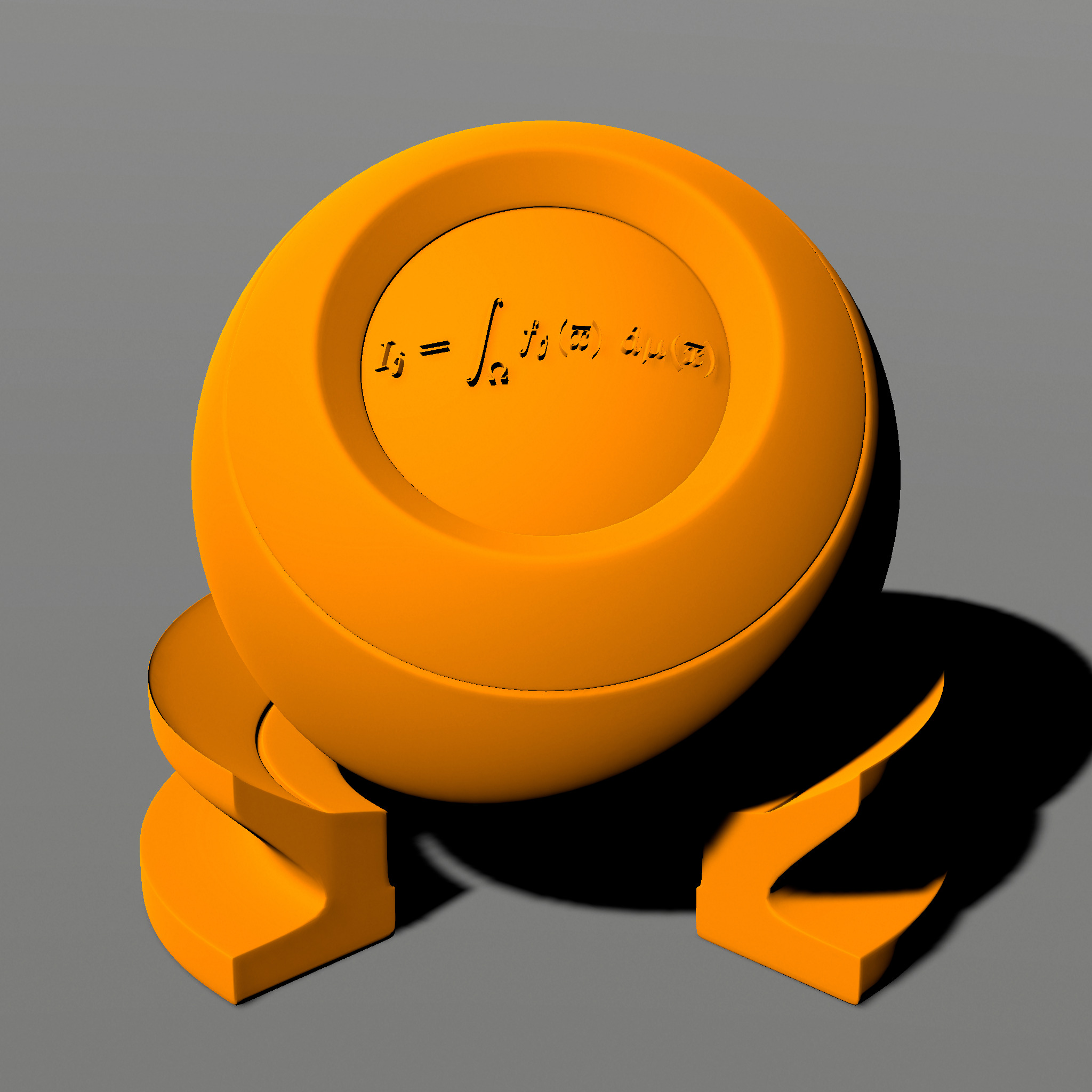}};
        \draw (spectral.south west) rectangle (spectral.north east);
        \node[below=3pt] at (spectral.south) {\textbf{(b) Unity rendering}};
    \end{tikzpicture}
    \vspace{-20pt}
    \caption{
        \textbf{Rendering in a Game Engine.} Our formulation is easily integrated into existing RGB rendering pipelines, and is well suited for (b) real-time rendering.
        We used Unity's Lit material, which implements the already available Disney Diffuse material~\protect\cite{Burley12} and not a strict Lambertian.
        Hence it does not fully match the (a) path-traced reference \edit{of the HERPIOYE material}.
        \label{fig:results_unity}
        \vspace{-10pt}
    }
\end{figure}

Figure~\ref{fig:results_unity} shows the rendering of a fluorescent material using both a reference path tracer with single scattering only and our implementation in Unity.
We show that our method produces a similar appearance.
We could not fully match the reference rendering since only Disney's Diffuse model~\protect\cite{Burley12} is implemented in Unity HDRP, not a Lambertian model.
Fully matching it would require rewriting all custom preintegration code.
Figure~\ref{fig:teaser} shows the impact of fluorescence when changing the lighting environment.

\subsection{Results in a Path-Tracer} \label{sec:results_path_tracing}
We implemented a backward path tracer with next-event estimation using the adjoint formulation. We compare with a fully stochastic spectral path tracer capable of handling fluorescence by following Mojzik~\etal~\cite{Mojzik18}. Our path tracer works in XYZ but can take arbitrary basis functions for its transport allowing us to add a 4$^{th}$ component for the UV band.

In Figure~\ref{fig:optimized_reradiation_matrix}, we compare our approach with the stochastic reference using a D65 illuminant. The D65 spectrum expands outside the visible range and the chosen fluorescent material partially reradiates UV light in the visible range. We show in Figure~\ref{fig:adding_uv_channels} that we can add a 4$^{th}$ UV-componnent when light contains a large amount of energy outside the visible range.

\begin{figure*}[p!]
    \hspace{-1.5cm}
    \begin{tikzpicture}[font=\footnotesize]
        \begin{scope}
            \pgfplotstableread[col sep = comma]{figures/scene/spectra/illuminant/350nm.csv}\mydata
            \begin{scope}[xshift=-0.5cm, yshift=-1.98cm, rotate=90]
                \node[rotate=90, anchor=center] at (2.0,1.7) {Gaussian at $350$nm};
                \draw[opacity=0.5] (-2.2,-0.4) -- (-2.2,0.7) -- (6,0.7) -- (6,-0.4);
                \begin{axis}[enlargelimits=false,width=5.58cm,height=3cm,grid=both,axis background/.style={fill=white},ymin=-0.01,ymax=1.5,xticklabel=\empty, yticklabel=\empty]
                    \addplot[thick,color=tab_blue,forget plot,] table[x index = {0}, y index = {1}]{\mydata};
                \end{axis}
                \node[rotate=90, anchor=center] at (-0.1,-0.3) {\textsc{IXCATAN}};
                \node[rotate=90, anchor=center] at (+3.9,-0.3) {\textsc{TEXTYELL}};
            \end{scope}
            \begin{scope}[xshift=0]
                \node[draw, line width=1pt, anchor=south west, inner sep=0] at (0,0) {\includegraphics[width=4cm]{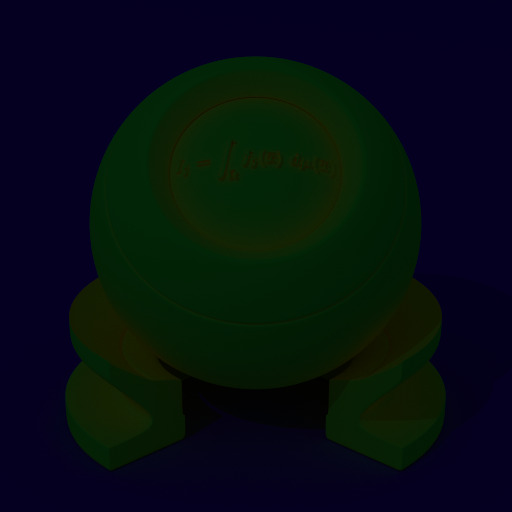}};
                \node[anchor=center] at (2,4.2) {\textbf{(a) Naive reduction}};
            \end{scope}
            \begin{scope}[xshift=4.2cm]
                \node[draw, line width=1pt, anchor=south west, inner sep=0] at (0,0) {\includegraphics[width=4cm]{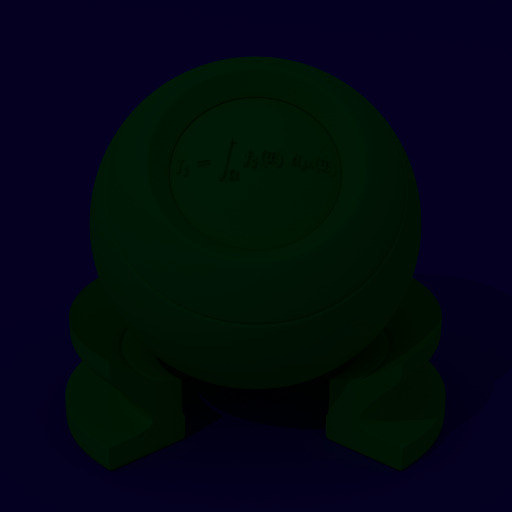}};
                \node[anchor=center] at (2,4.2) {\textbf{(b) Our reduction XYZ}};
            \end{scope}
            \begin{scope}[xshift=8.4cm]
                \node[draw, line width=1pt, anchor=south west, inner sep=0] at (0,0) {\includegraphics[width=4cm]{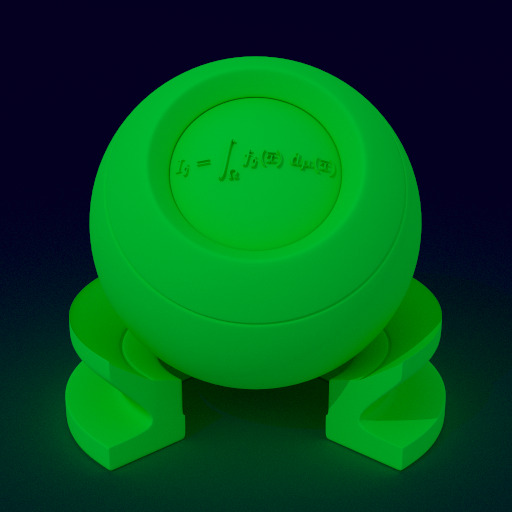}};
                \node[anchor=center] at (2,4.2) {\textbf{(c) Our reduction XYZU}};
            \end{scope}
            \begin{scope}[xshift=12.6cm]
                \node[draw, line width=1pt, anchor=south west, inner sep=0] at (0,0) {\includegraphics[width=4cm]{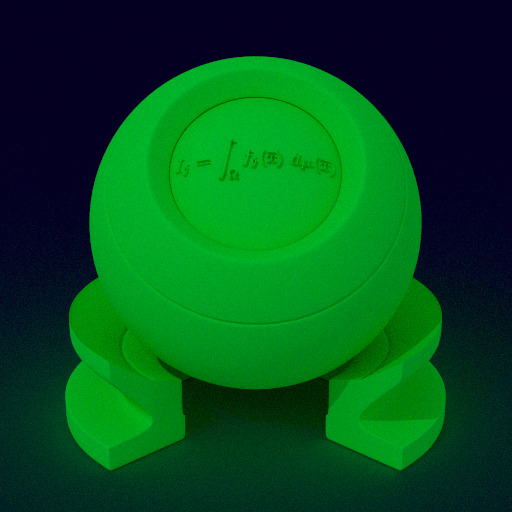}};
                \node[anchor=center] at (2,4.2) {\textbf{(r) Spectral reference}};
            \end{scope}
        \end{scope}

        \begin{scope}[yshift=-4.2cm]
            \begin{scope}[xshift=0cm]
                \node[draw, line width=1pt, anchor=south west, inner sep=0] at (0,0) {\includegraphics[width=4cm]{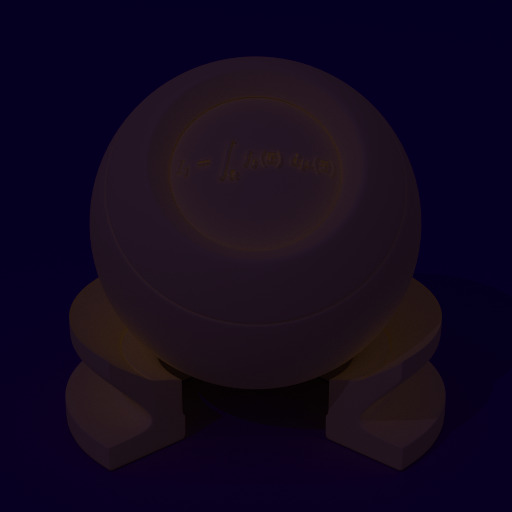}};
            \end{scope}
            \begin{scope}[xshift=4.2cm]
                \node[draw, line width=1pt, anchor=south west, inner sep=0] at (0,0) {\includegraphics[width=4cm]{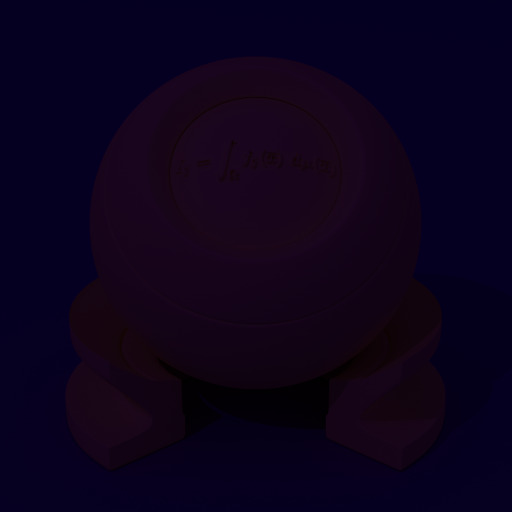}};
            \end{scope}
            \begin{scope}[xshift=8.4cm]
                \node[draw, line width=1pt, anchor=south west, inner sep=0] at (0,0) {\includegraphics[width=4cm]{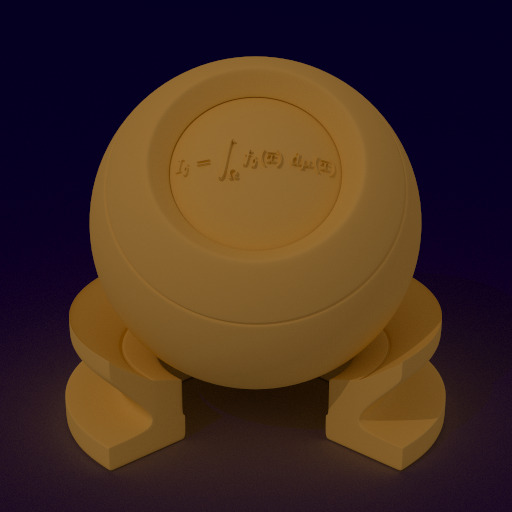}};
            \end{scope}
            \begin{scope}[xshift=12.6cm]
                \node[draw, line width=1pt, anchor=south west, inner sep=0] at (0,0) {\includegraphics[width=4cm]{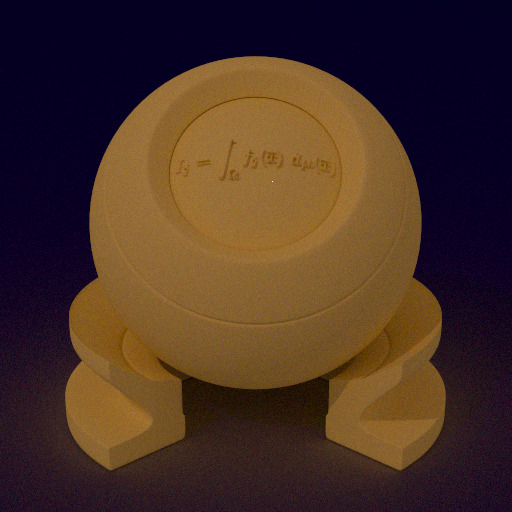}};
            \end{scope}
        \end{scope}

        \begin{scope}[yshift=-8.4cm]
            \pgfplotstableread[col sep = comma]{figures/scene/spectra/illuminant/450nm.csv}\mydata
            \begin{scope}[xshift=-0.5cm, rotate=90]
                \node[rotate=90, anchor=center] at (2.0,1.7) {Gaussian at $450$nm};
                \begin{axis}[enlargelimits=false,width=5.58cm,height=3cm,grid=both,axis background/.style={fill=white},ymin=-0.01,ymax=1.5,xticklabel=\empty, yticklabel=\empty]
                    \addplot[thick,color=tab_blue,forget plot,] table[x index = {0}, y index = {1}]{\mydata};
                \end{axis}
                \node[rotate=90, anchor=center] at (2.0,-0.3) {\textsc{IXCATAN}};
            \end{scope}
            \begin{scope}[xshift=0]
                \node[draw, line width=1pt, anchor=south west, inner sep=0] at (0,0) {\includegraphics[width=4cm]{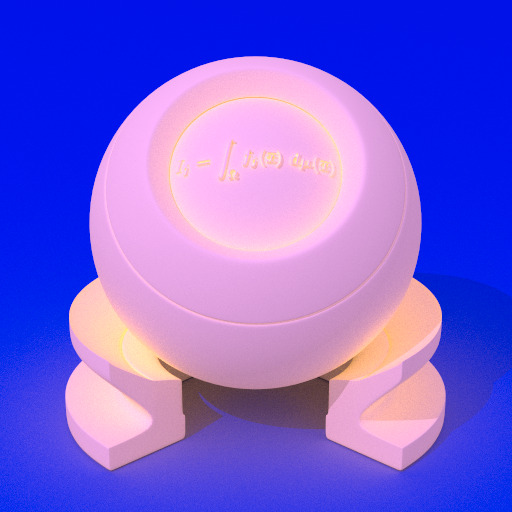}};
            \end{scope}
            \begin{scope}[xshift=4.2cm]
                \node[draw, line width=1pt, anchor=south west, inner sep=0] at (0,0) {\includegraphics[width=4cm]{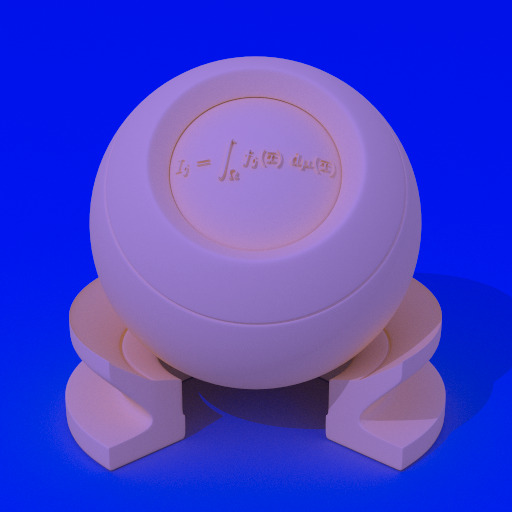}};
            \end{scope}
            \begin{scope}[xshift=8.4cm]
                \node[draw, line width=1pt, anchor=south west, inner sep=0] at (0,0) {\includegraphics[width=4cm]{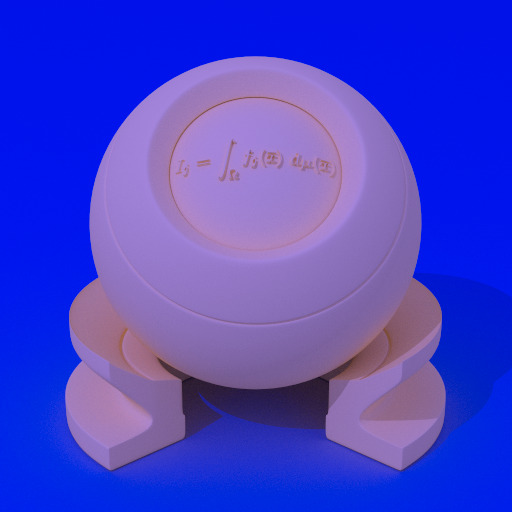}};
            \end{scope}
            \begin{scope}[xshift=12.6cm]
                \node[draw, line width=1pt, anchor=south west, inner sep=0] at (0,0) {\includegraphics[width=4cm]{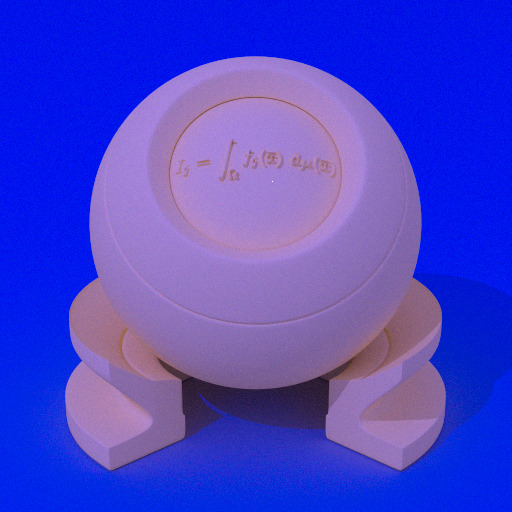}};
            \end{scope}
        \end{scope}

        \begin{scope}[yshift=-12.6cm]
            \pgfplotstableread[col sep = comma]{figures/scene/spectra/illuminant/D65.csv}\mydata
            \begin{scope}[xshift=-0.5cm, yshift=-1.98cm, rotate=90]
                \node[rotate=90, anchor=center] at (2.0,1.7) {\textsc{D65}};
                \draw[opacity=0.5] (-2.2,-0.4) -- (-2.2,0.7) -- (6,0.7) -- (6,-0.4);
                \begin{axis}[enlargelimits=false,width=5.58cm,height=3cm,grid=both,axis background/.style={fill=white},ymin=-0.01,ymax=150.,xticklabel=\empty, yticklabel=\empty]
                    \addplot[thick,color=tab_blue,forget plot,] table[x index = {0}, y index = {1}]{\mydata};
                \end{axis}
                \node[rotate=90, anchor=center] at (-0.1,-0.3) {\textsc{IXCATAN}};
                \node[rotate=90, anchor=center] at (+3.9,-0.3) {\textsc{TEXTYELL}};
            \end{scope}
            \begin{scope}[xshift=0]
                \node[draw, line width=1pt, anchor=south west, inner sep=0] at (0,0) {\includegraphics[width=4cm]{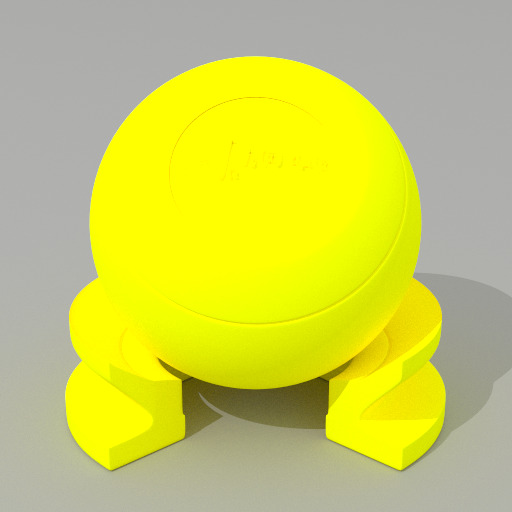}};
            \end{scope}
            \begin{scope}[xshift=4.2cm]
                \node[draw, line width=1pt, anchor=south west, inner sep=0] at (0,0) {\includegraphics[width=4cm]{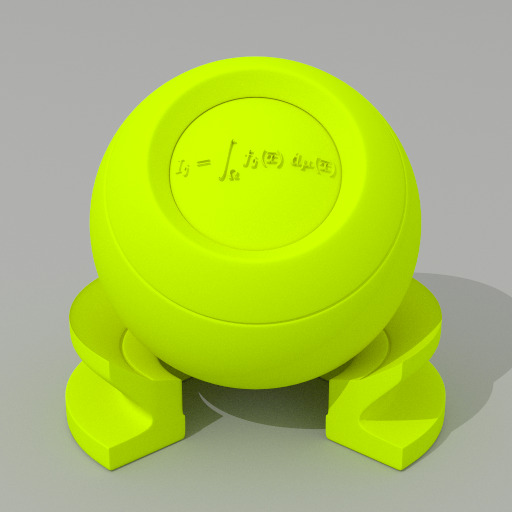}};
            \end{scope}
            \begin{scope}[xshift=8.4cm]
                \node[draw, line width=1pt, anchor=south west, inner sep=0] at (0,0) {\includegraphics[width=4cm]{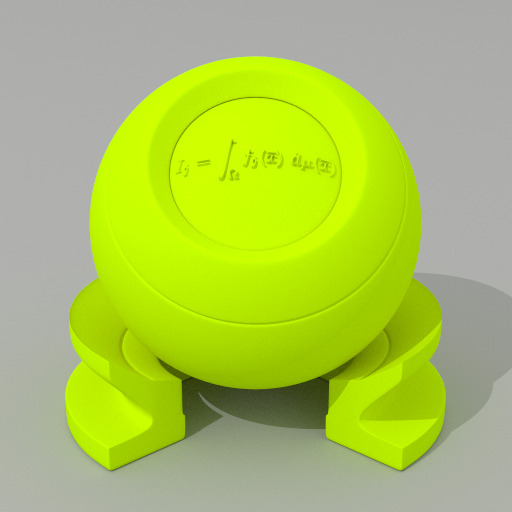}};
            \end{scope}
            \begin{scope}[xshift=12.6cm]
                \node[draw, line width=1pt, anchor=south west, inner sep=0] at (0,0) {\includegraphics[width=4cm]{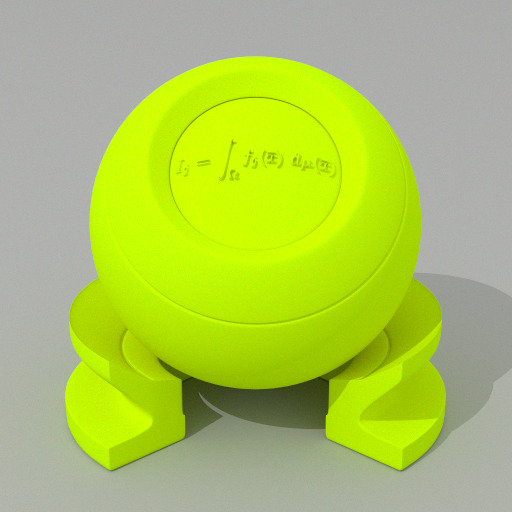}};
            \end{scope}
        \end{scope}

        \begin{scope}[yshift=-16.8cm]
            \begin{scope}[xshift=0]
                \node[draw, line width=1pt, anchor=south west, inner sep=0] at (0,0) {\includegraphics[width=4cm]{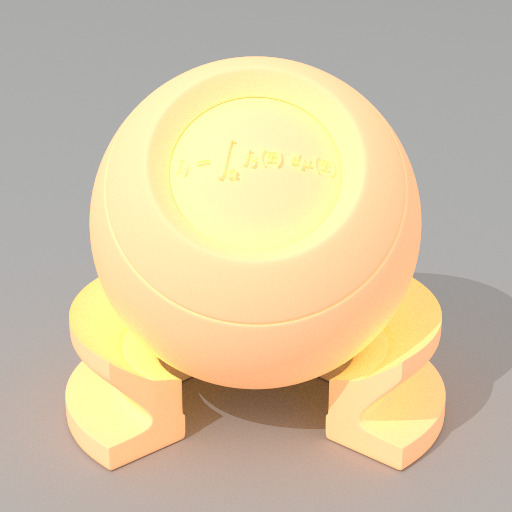}};
            \end{scope}
            \begin{scope}[xshift=4.2cm]
                \node[draw, line width=1pt, anchor=south west, inner sep=0] at (0,0) {\includegraphics[width=4cm]{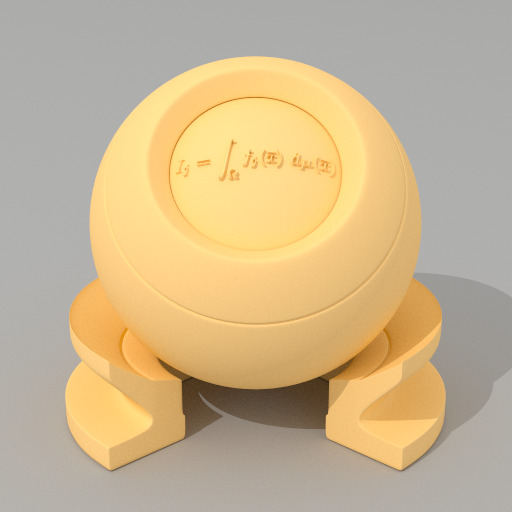}};
            \end{scope}
            \begin{scope}[xshift=8.4cm]
                \node[draw, line width=1pt, anchor=south west, inner sep=0] at (0,0) {\includegraphics[width=4cm]{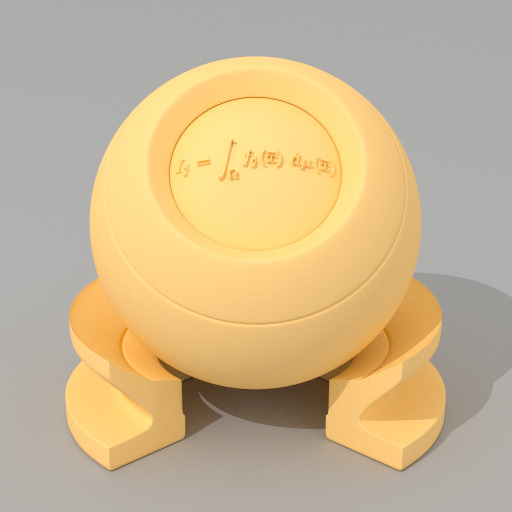}};
            \end{scope}
            \begin{scope}[xshift=12.6cm]
                \node[draw, line width=1pt, anchor=south west, inner sep=0] at (0,0) {\includegraphics[width=4cm]{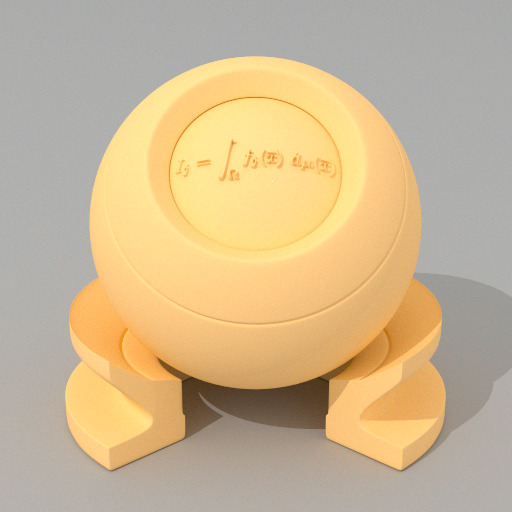}};
            \end{scope}
        \end{scope}
    \end{tikzpicture}
    \vspace{-0pt}
    \caption{
        \textbf{Validation in a path tracer.} We validate our method for different illuminants (left column) and different materials applied to a material probe. We display the method of (a) \edit{Naive}, (b) our XYZ method, (c) our XYZU method, and (r) the reference.
        \label{fig:validation}
    }
\end{figure*}

We show our technique works with a wide range of fluorescent materials in Figure~\ref{fig:validation}. It handles well multiple bounces with subtle color shifting as shown in the creases. The slight luminance decrease noted in Figure~\ref{fig:optimized_reradiation_matrix} is also noticeable here in XYZ for D65 and adding a UV band mitigates this issue. We further complement this set of results in our supplemental material.

\section{Discussion}
\label{sec-discussion}

\subsection{Limitations}

\begin{figure}[t]
    \begin{tikzpicture}[font=\footnotesize]
        \tikzmath{
            \margins = 6;
            \widthImage = (\the\linewidth - 2*\margins) / 3;
            \xImage1 = 0;
            \xImage2 = \xImage1 + \widthImage + \margins;
            \xImage3 = \xImage2 + \widthImage + \margins;
            \heightPlot = 41/49 * \widthImage;
        }
        \node[inner sep=0, anchor=west] (matrix_4) at (\xImage1 pt, 0) {\includegraphics[width=\widthImage pt]{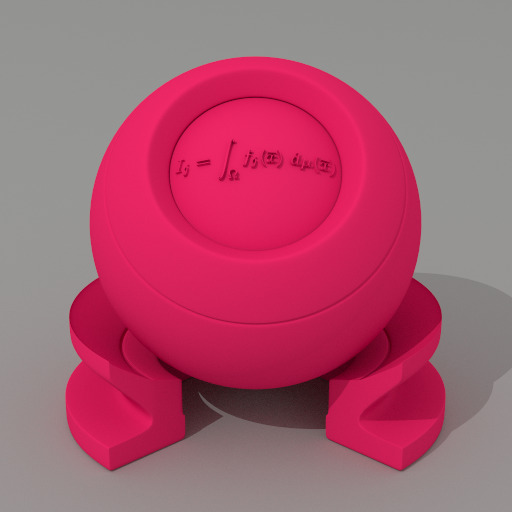}};
        \draw (matrix_4.south west) rectangle (matrix_4.north east);

        \node[inner sep=0, anchor=north, yshift=-\margins pt] (matrix_4_swipe) at (matrix_4.south) {\includegraphics[width=\widthImage pt, height=12pt]{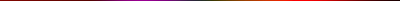}};
        \draw (matrix_4_swipe.south west) rectangle (matrix_4_swipe.north east);

        \pgfplotstableread[col sep=comma]{figures/scene/cmf/xyzu.csv}\mydata
        \begin{pgfinterruptboundingbox}
            \begin{axis}[
                at=(matrix_4.north),
                clip=true,
                anchor=south,
                yshift=\margins pt,
                width=\widthImage pt,
                height=\heightPlot pt,
                scale only axis,
                enlargelimits=false,
                grid=both,
                xmin=300,
                xmax=800,
                ymin=-0.01,
                ymax=2,
                xticklabel=\empty,
                yticklabel=\empty]
                \addplot[thick,color=tab_red  ,forget plot,] table[x index = {0}, y index = {1}]{\mydata};
                \addplot[thick,color=tab_green,forget plot,] table[x index = {0}, y index = {2}]{\mydata};
                \addplot[thick,color=tab_blue ,forget plot,] table[x index = {0}, y index = {3}]{\mydata};
                \addplot[thick,color=tab_pink ,forget plot,] table[x index = {0}, y index = {4}]{\mydata};
            \end{axis}
        \end{pgfinterruptboundingbox}

        \node[below=3pt] at (matrix_4_swipe.south) {\textbf{(a) Ours (4 bands)}};

        \node[inner sep=0, anchor=west] (matrix_7) at (\xImage2 pt, 0) {\includegraphics[width=\widthImage pt]{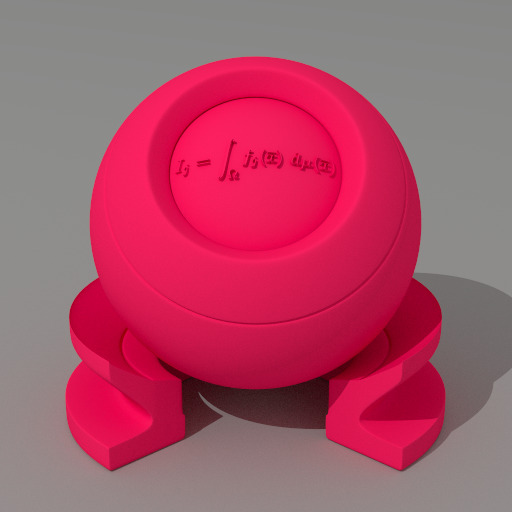}};
        \draw (matrix_7.south west) rectangle (matrix_7.north east);

        \node[inner sep=0, anchor=north, yshift=-\margins pt] (matrix_7_swipe) at (matrix_7.south) {\includegraphics[width=\widthImage pt, height=12pt]{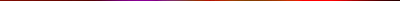}};
        \draw (matrix_7_swipe.south west) rectangle (matrix_7_swipe.north east);

        \pgfplotstableread[col sep=comma]{figures/scene/cmf/cmf_7.csv}\mydata
        \begin{pgfinterruptboundingbox}
            \begin{axis}[
                at=(matrix_7.north),
                anchor=south,
                yshift=\margins pt,
                width=\widthImage pt,
                height=\heightPlot pt,
                scale only axis,
                enlargelimits=false,
                grid=both,
                xmin=300,
                xmax=800,
                ymin=-0.01,
                ymax=2,
                xticklabel=\empty,
                yticklabel=\empty]
                \addplot[thick, color=tab_red!90!black, forget plot] table[x index = {0}, y index = {1}]{\mydata};
                \addplot[thick, color=tab_red!70!black, forget plot] table[x index = {0}, y index = {2}]{\mydata};
                \addplot[thick, color=tab_red!50!black, forget plot] table[x index = {0}, y index = {3}]{\mydata};
                \addplot[thick, color=tab_green!90!black, forget plot] table[x index = {0}, y index = {4}]{\mydata};
                \addplot[thick, color=tab_green!50!black, forget plot] table[x index = {0}, y index = {5}]{\mydata};
                \addplot[thick, color=tab_blue, forget plot] table[x index = {0}, y index = {6}]{\mydata};
                \addplot[thick, color=tab_pink, forget plot] table[x index = {0}, y index = {7}]{\mydata};
            \end{axis}
        \end{pgfinterruptboundingbox}

        \node[below=3pt] at (matrix_7_swipe.south) {\textbf{(b) Ours (7 bands)}};

        \node[inner sep=0, anchor=west] (spectral) at (\xImage3 pt, 0) {\includegraphics[width=\widthImage pt]{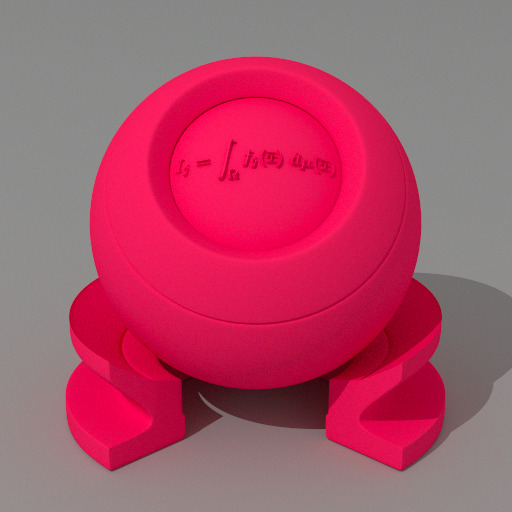}};
        \draw (spectral.south west) rectangle (spectral.north east);

        \node[inner sep=0, anchor=north, yshift=-\margins pt] (spectral_swipe) at (spectral.south) {\includegraphics[width=\widthImage pt, height=12pt]{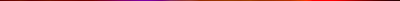}};
        \draw (spectral_swipe.south west) rectangle (spectral_swipe.north east);

        \node[inner sep=0, anchor=south, yshift=\margins pt] (spectral_rerad) at (spectral.north) {\includegraphics[width=\widthImage pt, height=\heightPlot pt]{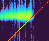}};
        \draw (spectral_rerad.south west) rectangle (spectral_rerad.north east);

        \node[below=3pt] at (spectral_swipe.south) {\textbf{(c) Spectral Reference}};
    \end{tikzpicture}
    \vspace{-19pt}
    \caption{
        {\textbf{Limitation.}}
        The use of $4$ bases in (a) is not sufficient to reproduce the spectral rendering of the HERPIMAG material, lit with a D65 illuminant in (c).
        When looking at the color swipes (bottom row), we observe a part of the reradiation is missing around $450\:\mathrm{nm}$. This can be addressed by increasing the number of bases.
        In (b) we split the CIE-XYZ 2006 2$^\circ$ CMFs (as detailed in Appendix~\ref{sec:alternate-bases}) to better capture color shifting effects while keeping the reconstruction of the XYZ color linear.
        \label{fig:failure_test}
        \vspace{-15pt}
    }
\end{figure}

\paragraph*{Spectral manifold.} Our method can only faithfully reproduce light transport within the manifold of the basis functions. Hence, in some cases, our method could not reproduce the same color as spectral rendering.
We show in Figure~\ref{fig:failure_test}~(a) a material where our method fails to correctly capture the appearance.

\paragraph*{Number of bases.} However, we can improve the quality of the approximation by using more basis functions.
This is at the expense of a higher computational cost.
For example, we show in Figure~\ref{fig:failure_test}~(b) {using $7$ bases by splitting the X and Y channels (as explained in Appendix~\ref{sec:alternate-bases})} improves the quality of our approximation. Note however that for backward path tracing it requires to track $7 \times 7$ matrices, and use Equation~\ref{eq:refinedT} for $T$.

\subsection{Future work}
\paragraph*{Alternate basis.} So far, we have experimented with non-orthogonal bases derived from XYZ for ease of reconstruction. However, our method is generic enough to allow any form of basis. It would be interesting to see if less overlapping bases would yield better results.
Working in the specific color spaces of cameras' (or animals') sensitivity functions might also be a sensible application-dependent choice.

\edit{\paragraph*{Color transformation.} We currently determine $T$ by chaining the upsampling operator of the transport basis $\mathbf{b}$ and the downsampling operator of the sensor's basis $\mathbf{s}$ with $T = S^{\top}_\mathbf{s} \times \tilde{S}_\mathbf{b}$. This is a valid solution but may not be optimal for human vision. Alternatively, we could optimize $T$ as done for color space transformations~\cite{Langlands20}. Using perceptual distances such as $\Delta E^*_{2000}$ might provide a better result compared to our linear approach.
}

\paragraph*{Glossy fluorescent materials.} We have worked with Lambertian fluorescent models for which measured materials are easily available. Our method could apply to the work of Hullin~\etal~\cite{Hullin10} since their decomposition of fluorescent materials is linear. For the model of Benamira and Pattanaik~\cite{Benamira23}, more work is needed as the fluorescence is dependent on the incident angle.

\paragraph*{Other phenomena.} We wonder if our formulation of reduced light transport based on downsampling/upsampling could be applied to other spectral phenomena known to require spectral rendering, such as atmospheric/water rendering~\cite{Elek10}. Provided that directional and spectral dimensions are independent, our formulation remains applicable.

\paragraph*{Measuring Fluorescence.} Our solution might be used to design simpler fluorescence measurement setups without dense spectral control on the source and sensor, provided one is only interested in capturing the \emph{reduced} fluorescence matrix.

\subsection{Conclusion}
We have presented a new formulation to reproduce fluorescence appearance in non-spectral renderers such as RGB rasterizers or path tracers. Our method provides a principled way to reduce the dense reradiation matrix of BRRDF materials to enable correct rendering of fluorescence. Our method permits to visually reproduce a spectral reference as well as to incorporate non-visible light leaking at a small cost in non-spectral rendering.



\printbibliography

\appendix
\section{Alternative bases}
\label{sec:alternate-bases}

We have explored a couple choices of alternative basis functions besides the XYZ CMFs.

First, we have considered adding a single basis $U$ to capture reradiation from the ultraviolet, focusing on the UVA range (\ie, $[315-400]\:\mathrm{nm}$) which accounts for approximately $95$\% of the ultraviolet radiation reaching the Earth's surface.
To this end, we build a Partition of Unity (PU) over the $[300-800]\:\mathrm{nm}$ range, using $5$ second-order B-Spline bases.
We obtained satisfying results with the first of these B-Splines as our additional UV basis $U$: it is monotonically-decreasing, {with $U>0.5$ for wavelengths below $400\:\mathrm{nm}$.}
Yet other options could be considered, for instance when knowing specifics of material reradiation in the UV range.

Second, we have experimented with the use of more than $3$ bases in the visual range.
Our approach is to leave the $Z$ CMF basis unchanged, but split the large $Y$ CMF basis in two, and the bimodal $X$ CMF basis in three.
Starting with the $Y$ basis, we split the corresponding CMF to get $\bar{y} = \bar{y}_1 + \bar{y}_2$, with:
\begin{align}
    \bar{y}_1(\lambda) & = \bar{y}(\lambda) \mathcal{S}(\lambda; \mu_y, \sigma_y) \\
    \bar{y}_2(\lambda) & = \bar{y}(\lambda) [1-\mathcal{S}(\lambda; \mu_y, \sigma_y)],
\end{align}
where $\mathcal{S}$ is a smoothstep function centered on $\mu_y$ and scaled by $\sigma_y$.
It is given by $\mathcal{S}(\lambda; \mu_y, \sigma_y) = 3 \lambda_y^2 - 2 \lambda_y^3$ with $\lambda_y = \frac{1}{2} \left[\frac{\lambda-\mu_y}{\sigma_y} + 1 \right]_0^1$ clipped to the $[0,1]$ range.
{We use $\mu_y=570\:\mathrm{nm}$ and $\sigma_y=60\:\mathrm{nm}$.}

The splitting of the $X$ basis is similar, except we separate it in three $\bar{x} = \bar{x}_1 + \bar{x}_2 + \bar{x}_3$, with $\bar{x}_1$ (resp. $\bar{x}_2 + \bar{x}_3$) corresponding to the mode below (resp. above) $500$nm. Formally, we have:
\begin{align}
    \bar{x}_1(\lambda) & = \bar{x}(\lambda) [1-\mathcal{S}(\lambda; 500, 2)] \\
    \bar{x}_2(\lambda) & = \bar{x}(\lambda) \mathcal{S}(\lambda; 500, 2) \mathcal{S}(\lambda, \mu_x, \sigma_x) \\
    \bar{x}_3(\lambda) & = \bar{x}(\lambda) \mathcal{S}(\lambda; 500, 2) [1-\mathcal{S}(\lambda, \mu_x, \sigma_x)].
\end{align}
{We use $\mu_x=590\:\mathrm{nm}$ and $\sigma_x=60\:\mathrm{nm}$.}

Adding the $U$ basis for ultraviolet reradiation yields $K=7$ bases.
We obtain the final color by multiplying $\mathbf{c}_o$ by:
\begin{align}
    \label{eq:refinedT}
    T = \left[
        \begin{array}{ccccccc}
            1 & 1 & 1 & 0 & 0 & 0 & 0 \\
            0 & 0 & 0 & 1 & 1 & 0 & 0 \\
            0 & 0 & 0 & 0 & 0 & 1 & 0 \\
            0 & 0 & 0 & 0 & 0 & 0 & 1
        \end{array}
    \right].
\end{align}
The advantage of splitting XYZ basis functions to get finer bases is that they map to the final XYZ color through $T$ without incurring further approximations.
In general, we suggest that any set of basis functions that may be linearly combined (through $T$) to form the XYZ CMFs is an adequate choice.

{ Note that when rendering with forward path tracing, this requires transporting $7 \times 7$ reduced matrices denoted $\reduxo_7$.
An experiment we leave to future work would be to transport $4 \times 4$ reduced matrices $\reduxo_4$, but use a $4 \times 7$ matrix $\reduxo_{7 \rightarrow 4}$ when connecting with the light source.
The latter matrix is then given by $\reduxo_{7 \rightarrow 4} = T \times \reduxo_7$.}

\end{document}